\documentclass[prl,aps,twocolumn,superscriptaddress,nofootinbib,longbibliography]{revtex4-2}
\usepackage{graphicx} 
\usepackage{dcolumn}  
\usepackage{bm}       
\usepackage{bbm}       
\usepackage{amsmath}
\usepackage{epsfig}
\usepackage{color}
\usepackage{ulem}
\usepackage[colorlinks,linkcolor=black,citecolor=blue,urlcolor=blue,hyperindex,driverfallback=dvipdfm]{hyperref}
\usepackage{orcidlink}

\begin{document}
\title{Chiral Light-Matter Interactions with Thermal Magnetoplasmons in Graphene Nanodisks}

\author{Mikkel Have Eriksen\,\orcidlink{0000-0002-0159-0896}}
\email{meri@mci.sdu.dk}
\affiliation{POLIMA---Center for Polariton-driven Light--Matter Interactions, University of Southern Denmark, Campusvej 55, DK-5230 Odense M, Denmark}

\author{Juan R. Deop-Ruano\,\orcidlink{0000-0002-2613-5873}}
\affiliation{Instituto de Qu\'imica F\'isica Blas Cabrera (IQF), CSIC, 28006 Madrid, Spain}

\author{Joel D. Cox\,\orcidlink{0000-0002-5954-6038}}
\email{cox@mci.sdu.dk}
\affiliation{POLIMA---Center for Polariton-driven Light--Matter Interactions, University of Southern Denmark, Campusvej 55, DK-5230 Odense M, Denmark}
\affiliation{Danish Institute for Advanced Study, University of Southern Denmark, Campusvej 55, DK-5230 Odense M, Denmark}

\author{Alejandro Manjavacas\,\orcidlink{0000-0002-2379-1242}}
\email{a.manjavacas@csic.es}
\affiliation{Instituto de Qu\'imica F\'isica Blas Cabrera (IQF), CSIC, 28006 Madrid, Spain}

\date{\today}

\begin{abstract}
We investigate the emergence of self-hybridized thermal magnetoplasmons in doped graphene nanodisks at finite temperatures when subjected to an external magnetic field. Using a semianalytical approach, which fully describes the eigenmodes and polarizability of the graphene nanodisks, we show that the hybridization originates from the coupling of transitions between thermally populated Landau levels and localized magnetoplasmon resonances of the nanodisks. Owing to their origin, these modes combine the extraordinary magneto-optical response of graphene with the strong field enhancement of plasmons, making them an ideal tool for achieving strong chiral light-matter interactions, with the additional advantage of being tunable through carrier concentration, magnetic field, and temperature. As a demonstration of their capabilities, we show that the thermal magnetoplasmons supported by an array of graphene nanodisks enable chiral perfect absorption and chiral thermal emission.
\end{abstract}
\maketitle

A photonic structure is considered chiral if it cannot be superimposed onto its mirror image through translations and rotations \cite{wagniere2007on}. Chiral photonic structures respond differently to the left- (LCP) and right-handed (RCP) circularly polarized  components of the electromagnetic field, resulting in a phenomenon known as circular dichroism \cite{born1999principles,mun2020electromagnetic}. This ability makes them ideal tools for the detection, identification, and manipulation of chiral molecules \cite{garcia2018enantiomer,paiva2020chiral,warning2021nanophotonic,kim2022enantioselective}, which is a problem of paramount importance in biotechnology \cite{nguyen06chiral}. Chiral photonic structures can also be employed to control and modify the polarization state of light \cite{collins2017chirality,mun2020electromagnetic,avalos2022chiral} and to generate circularly polarized thermal radiation \cite{lee2007circularly,wadsworth2011broadband,dyakov2018magnetic,wang2023observation}.  These prospects have driven an intense research effort into developing different chiral photonic elements, including, among others, emitters \cite{govorov2010theory,schaferling2012tailoring,stamatopoulou2022reconfigurable}, cavities \cite{voronin2022single,baranov2023toward}, absorbers \cite{ouyang2018near,wu2020chiral}, metasurfaces \cite{wu2018high,deng2024advances}, and collective modes in periodic arrays \cite{goerlitzer2020chiral,movsesyan2022engineering,cerdan2023chiral}. 

\begin{figure*}
\begin{center}
\includegraphics[width=\linewidth]{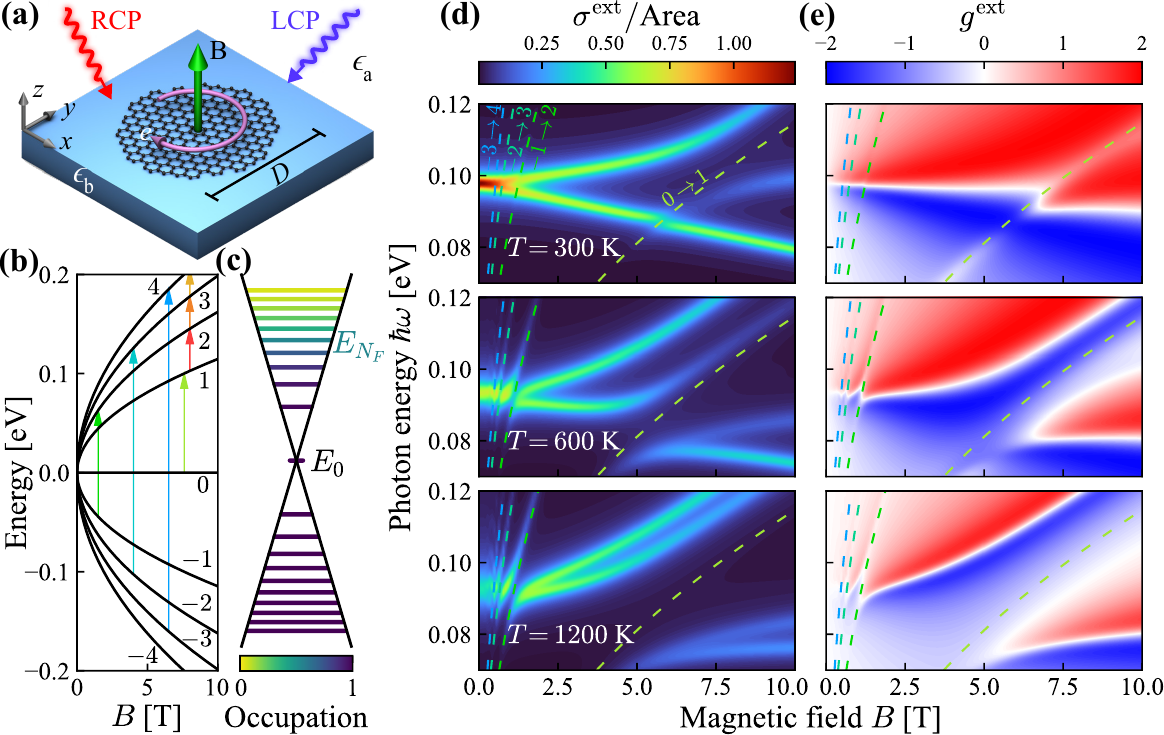}
\caption{{\bf Magneto-optical graphene nanodisk and thermal magnetoplasmons.} (a) Schematic of a graphene disk with diameter $D$, subjected to a perpendicular static magnetic field and excited by incoming LCP and RCP light. (b) First four LLs plotted as a function of the magnetic field. The vertical arrows indicate some of the possible transitions between these levels. (c) Conical band-structure of graphene split into LLs, which are color-coded to indicate the electron population at $T=300\,$K for $B=5\,$T and $E_{\rm F}=0.2\,$eV. (d) Extinction cross-section of a graphene nanodisk in vacuum for linearly polarized light as a function of $B$. We assume $D=120\,$nm, $E_{\rm F}=0.2\,$eV, and consider three different temperatures: $T=300\,$K, $T=600\,$K, and $T=1200\,$K. 
(e) Extinction dissymmetry factor corresponding to the results of panel (d), calculated using Eq.~\eqref{eq:extinction_dissymmetry_factor}. The dashed lines in (d) and (e) signal the transitions between LLs and are color-coded to match the arrows in (b).}
\label{fig:fig1}
\end{center}
\end{figure*}

Another approach to achieving circular dichroism involves the use of materials with a strong magneto-optical response. Among these, graphene---a single layer of carbon atoms---stands out as one of the most promising platforms due to its unique electronic properties \cite{sounas2011electromagnetic, sounas2012gyrotropy}. In the presence of an external static magnetic field directed normally to a graphene sheet, the optical conductivity of this material exhibits off-diagonal Hall components that result in an extraordinary magneto-optical response, as exemplified by reports of giant Faraday rotation \cite{crassee2011giant,tymchenko2013faraday,han2023strong}. 

Meanwhile, when doped with charge carriers, graphene nanostructures are able to support localized surface plamons \cite{gonccalves2016introduction}. These excitations, whose strength and frequency are determined by the level of doping, can boost light-matter interaction due to the extraordinary near-field enhancement and cross-sections that they provide at terahertz and infrared frequencies \cite{garcia2014graphene,jadidi2016nonlinear,ni2016ultrafast,cox2018nonlinear}. A notable example is the possibility of achieving perfect absorption using an array of graphene nanostructures arranged in a Salisbury screen configuration---a concept initially predicted theoretically \cite{thongrattanasiri2012complete} and later confirmed experimentally  \cite{kim2018electronically}.

Due to the combined effects of charge carrier doping and applied static magnetic field, graphene nanostructures support magnetoplasmons that enable strong chiral light-matter interactions \cite{wang2011edge, yan2012infrared,wang2012excitation,kort2015active,jiao2019shape,tamagnone2016near}.  In principle, the characteristics of these excitations are largely determined by the doping level and the strength of the static magnetic field \cite{yan2012infrared,poumirol2017electrically}.  However, the linear band structure and two-dimensional (2D) nature of this material, which are at the origin of its high optoelectronic tunability, together with its low electronic heat capacity, facilitate substantial temperature-induced changes in the optical response of graphene nanostructures \cite{yu2017ultrafast, yu2018photothermal,cox2019single}. This characteristic presents new opportunities for controlling and manipulating graphene magnetoplasmons but demands exploring how the interplay of a static magnetic field, carrier doping, and temperature influences their behavior.

Here, we theoretically investigate new strategies to thermally control chiral light absorption and emission using graphene magnetoplasmons. We demonstrate that, at finite temperature and in the presence of a perpendicular static magnetic field, the magnetoplasmons supported by graphene nanodisks undergo strong hybridization with transitions between thermally populated Landau levels (LLs) that become more prominent at elevated electron temperatures, leading to the formation of hybrid thermal magnetoplasmons (TMPs). Our investigation is based on a semianalytical formalism that describes the eigenmodes and the polarizability of magneto-optical 2D nanodisks, which we introduce here to fully characterize the properties of TMPs in isolated nanodisks. By combining this formalism with a coupled dipole model and transfer matrix calculations, we demonstrate that the TMPs supported by arrays of graphene nanodisks enable chiral perfect absorption (i.e., perfect absorption for only one light handedness) when paired with a gold mirror in a Salisbury screen configuration. Furthermore, we demonstrate that the same system generates fully circularly polarized thermal radiation of a single handedness. These findings extend the known capability of graphene plasmons to mediate thermal  emission \cite{manjavacas2014graphene,brar2015electronic} and pave the way for developing novel nonreciprocal thermal emitters \cite{khandekar2020new,shi2022thermal,shayegan2023direct} capable of advancing solar energy harvesting technologies \cite{yang2024nonreciprocal}.

We begin by analyzing the optical response of an individual graphene nanodisk, as depicted in Fig.~\ref{fig:fig1}(a). The nanodisk, with diameter $D$, is located at the interface between two media with dielectric functions $\epsilon_{\rm a}$ and $\epsilon_{\rm b}$, and is subjected to an externally applied static magnetic field $B$.
To characterize its magneto-optical response, we employ a semianalytical approach described in detail in the Appendix. This approach is based on turning the self-consistent equation relating the potential and the induced charge in the nanodisk into an eigenvalue problem \cite{fetter1986magnetoplasmons,wang2012edge,garcia2014graphene,christensen2014classical,karanikolas2016tunable,karanikolas2017localized,yu2017analytical,muniz2020two}. Upon solving this problem and obtaining the eigenmodes characterizing the response of the nanodisk, we use them to construct its electrostatic polarizability. That polarizability is then corrected to account for the effect of retardation \cite{deop2022optical}. Importantly, the material properties of the structure enter the model only through the optical conductivity. Therefore, our semianalytical approach can be used to describe a magneto-optical nanodisk made from any material, such as doped semiconductors \cite{schimpf2014charge,yin2018plasmon} or noble metals \cite{manjavacas2014tunable,el-fattah2019plasmonics}, provided that the structure can be effectively considered two-dimensional.

For the specific case of graphene in the presence of a perpendicular static magnetic field, we adopt the well-established magneto-optical conductivity model from  Ref.~[\citenum{ferreira2011faraday}]. Briefly (see the Appendix for more details), the longitudinal $\sigma_{\parallel}$ and Hall $\sigma_{\perp}$ components of the conductivity tensor are obtained as sums over LLs with energies $E_n={\rm sgn}(n) \hbar v_{\rm F}\sqrt{2|n|}/ L_B$, defined by the graphene Fermi velocity $v_{\rm F} \approx c/300$ and the magnetic length $L_B=\sqrt{\hbar c/eB}$. In addition to the LL energies, the sums also depend on the LL occupation factors $f_n=[{\rm e}^{(E_n-\mu)/k_{\rm B} T}+1]^{-1}$, where $\mu$ represents the chemical potential, and on the electron damping rate $\tau^{-1}$. Throughout this work, we assume $\tau^{-1} = e v_{\rm F}^2 / \mu_{\rm DC} E_{\rm F}$ with a DC mobility $\mu_{\rm DC}=10^4\,$cm$^2$/Vs and Fermi energy $E_{\rm F}$. However, we also discuss the effect of temperature on $\tau^{-1}$ later.

The electron distribution governing the magnetoplasmonic response of a graphene nanodisk is the result of the interplay between carrier concentration, magnetic field, and temperature. As explained above and illustrated by Figs.~\ref{fig:fig1}(b) and \ref{fig:fig1}(c), the LL energies are determined by the magnetic field strength, while their occupation follows a Fermi-Dirac distribution governed by temperature and carrier concentration.  Consequently, for a given carrier concentration, the chemical potential, the temperature, and the magnetic field are intricately interdependent due to carrier conservation \cite{yu2017ultrafast}. This important phenomenon is incorporated in the theory and results presented in this work, as we explain in detail in the Appendix (see Fig.~\ref{fig:Chemical_pot}). There, we also provide the explicit expressions relating $\mu$, $T$, and $B$ with $E_{\rm F}$ [see Eqs.~\eqref{eq:Chemical_potential_BField_num_find} and \eqref{eq:Chemical_potential_BField_finiteT_num_find}],  as well as a discussion of the temperature dependence of the electron damping rate and the conductivity (see Figs.~\ref{fig:temp_dependent_scattering_rate} and \ref{fig:conductivity}).

Since the magnetoplasmons supported by the graphene nanodisk are strongly affected by the energy and population of the LLs, their properties inherit a dependence on carrier concentration, magnetic field, and temperature. Furthermore, the smearing out of the Fermi-Dirac distribution at finite temperature enables transitions between partially populated LLs, which can hybridize with the magnetoplasmons and lead to the emergence of new resonances. To explore these intriguing behaviors, we calculate the extinction cross-section of a magneto-optical graphene nanodisk across different temperatures. We assume that the nanodisk is excited with linearly-polarized light and has diameter $D=120\,$nm and Fermi energy $E_{\rm F} = 0.2\,$eV. The corresponding results are shown in Fig.~\ref{fig:fig1}(d), with the upper, middle, and lower plots corresponding, respectively, to $300\,$K, $600\,$K, and $1200\,$K. As demonstrated in Fig.~\ref{fig:ext_Comsol}, these results are in excellent agreement with direct numerical solutions of Maxwell’s equations obtained using a finite element method. Examining them, we observe that the presence of the magnetic field induces a splitting of the plasmon resonance into two branches. However, unlike the well-known magneto-optical response for $T=0\,$K \cite{wang2012edge}, which is analyzed in Fig.~\ref{fig:zero_temp}, the splitting for finite temperatures is asymmetric and the branches evolve differently as $B$ increases. The origin of this behavior is the coupling of the magnetoplasmon with specific LL transitions, which results in the emergence of new hybridized modes. This coupling manifests in the spectrum as anti-crossing features, located at the intersection between each of the magnetoplasmon branches and the different LL transitions. The latter are marked with dashed curves color-coded to match the arrows of Fig.~\ref{fig:fig1}(b). 

The thermally-induced self-hybridization of magnetoplasmons with LL transitions signals the emergence of TMPs. These excitations are reminiscent of the Landau-phonon polaritons observed in Ref.~[\citenum{wehmeier2024landau}], which originate from the hybridization of Dirac magnetoexciton modes of graphene with phonon polaritons of hexagonal boron nitride crystals, as well as the hybrid polaritons proposed in Ref.~[\citenum{tserkezis2024selfhybridisation}] that arise from the hybridization of interband transitions and Mie resonances. A closer examination of the results in Fig.~\ref{fig:fig1}(d) reveals that, for $T=300\,$K, the only signatures of TMPs in the extinction cross-section appear under strong magnetic fields ($B\gtrsim5\,$T). They correspond to the hybridization between the  magnetoplasmons and the $n=0\to n=1$ LL transition. Raising the temperature to $T=600\,$K and $T=1200\,$K increasingly unbalances the population of these two LLs, which enhances the transition rate. Consequently, the coupling of this transition with the magnetoplasmon is strengthened, resulting in a much more pronounced anti-crossing feature in the spectrum. A qualitatively similar dependence with temperature is observed for the TMP anti-crossings that emerge at low magnetic fields ($B\lesssim 1\,$T), which are associated with the interband LL transitions (i.e., from negative to positive $n$) indicated by the labels. The increase in carrier concentration, which is analyzed in Fig.~\ref{fig:ext_dissymmetry_factor}, shifts the magnetoplasmon to higher energies, thus forcing the TMPs to appear at stronger magnetic fields. Furthermore, due to the larger occupation of the LLs involved in these transitions, the appearance of strong anti-crossing features associated with the TMPs requires higher temperatures.

The excitation of TMPs strongly influences the chiral response of the nanodisks. To quantify this effect, we calculate the extinction dissymmetry factor in Fig.~\ref{fig:fig1}(e), defined as
\begin{equation} \label{eq:extinction_dissymmetry_factor}
g^{\rm ext} = 2\frac{\sigma_{+}^{\rm ext}-\sigma_{-}^{\rm ext}}{\sigma_{+}^{\rm ext}+\sigma_{-}^{\rm ext}}.
\end{equation}
Here, $\sigma_{\pm}^{\rm ext}=4\pi k{\rm Im}\{\hat{\bf e}_\pm^*\cdot\alpha\cdot\hat{\bf e}_\pm\}$ is the extinction cross-section for LCP ($+$) and RCP ($-$) light, while $\alpha$ represents the polarizability of the nanodisk and $\hat{\bf e}_\pm=(\hat{\bf x}\pm {\rm i}\hat{\bf y})/\sqrt{2}$. Analyzing the extinction dissymmetry factor, we find that the TMPs give rise to a strong chiral response, with the absolute value of $g^{\rm ext}$ reaching its maximum even for moderate values of $B$. Furthermore, the complex patterns in the extinction cross-section, arising from the TMP anti-crossings, are transferred to the maps of $g^{\rm ext}$. This results in a much richer response that can be tuned not only by the carrier concentration and magnetic field but also by the temperature.

As discussed above, the results of Figs.~\ref{fig:fig1}(d) and (e) are obtained using a constant value for the electron damping rate $\tau^{-1}$. However, $\tau^{-1}$ is expected to increase with temperature \cite{sohier2014phonon,yuan2020room,dias2020thermal}. To analyze how such temperature dependence impacts the response of the graphene nanodisk, we repeat the calculations of Fig.~\ref{fig:fig1}(d) for reduced values of $\mu_{\rm DC}$. The corresponding results, which are summarized in Fig.~\ref{fig:varying_loss}, reveal that, although the decrease in mobility causes the spectral features to blur, the hybridization behavior remains clearly visible even for the lowest mobility under consideration ($\mu_{\rm DC}=1000\,$cm$^2$/Vs).

While the TMPs of an isolated graphene nanodisk provide relatively strong chiral light-matter interactions, practical implementations require the use of an ensemble to accumulate a sufficiently strong response in the far field. Among various possible configurations, a periodic array enables the coherent enhancement of the response of the individual nanodisks \cite{deop2022optical}, while maintaining minimal structural complexity \cite{kim2018electronically}. Indeed, when integrated with a mirror and a spacer in a Salisbury screen configuration \cite{salisbury1952absorbent,alaee2017theory}, these systems can lead to perfect absorption \cite{thongrattanasiri2012complete}.  
In the following, we explore the possibility of combining these capabilities with the strong magneto-optical response of the TMPs discussed above to achieve chiral perfect absorption, controllable through carrier concentration, magnetic field, and temperature.
To study the response of such a system, we exploit the coupled dipole model, which produces very accurate results provided that the nanostructures are small compared to the wavelength of light and the lattice period \cite{garciadeabajo2007colloquium,kolkowski2019lattice,cuartero2020super}. Within this framework, the dipole moment induced in an arbitrary nanodisk is given by the sum of the contributions from the external field and the fields produced by the rest of the elements of the array (see the Appendix for a detailed derivation). Furthermore, the periodicity of the system ensures that the dipole can be expressed as a Bloch wave \cite{zundel2022lattice}, which under normal incidence excitation, reduces to ${\bf p}=\alpha^{\rm eff}\cdot{\bf E}^{\rm ext}$ with $\alpha^{\rm eff} = \left[\alpha^{-1} - \mathcal{G}\right]^{-1} \nonumber$
representing the effective polarizability of the array. This quantity encapsulates all the information about the optical response of the array, which arises both from the individual response of each nanodisk, described by $\alpha$, and from their interactions, captured by the lattice sum $\mathcal{G}$\cite{zundel2021lattice}. The latter is given by $\mathcal{G} = \sum_{j\neq 0} {\bf G}_{0j}$ in terms of the dyadic Green tensor for a homogeneous medium \cite{novotny2012principles,zundel2022lattice}.

\begin{figure}
\begin{center}
\includegraphics[width=\linewidth]{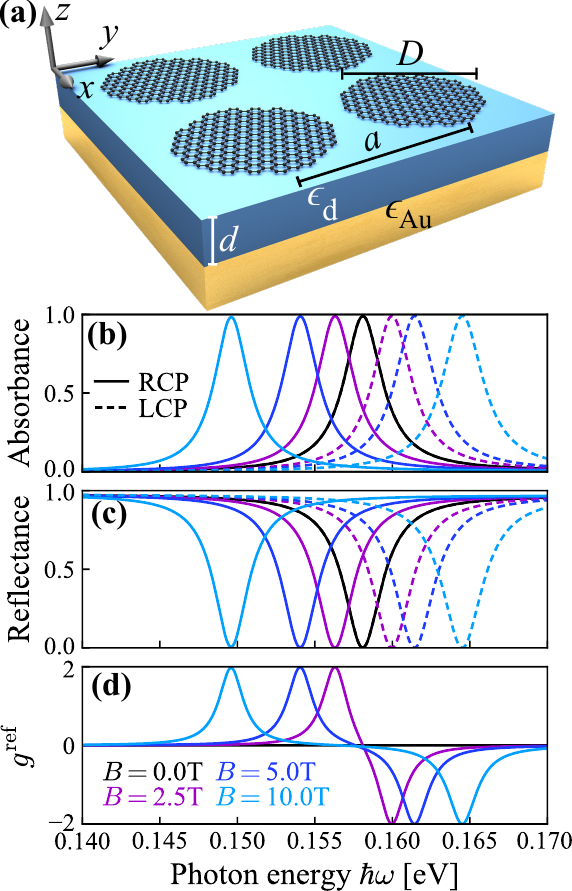}
\caption{{\bf Chiral perfect absorption.} (a) Schematic of the system under investigation. (b,c) Absorbance (b) and reflectance (c) at normal incidence for the different magnetic field strengths indicated in the legend. Dashed and solid curves correspond to LCP and RCP light, respectively. We consider an array with period $a=2D$ composed of nanodisks with $D=60\,$nm, $E_{\rm F}=0.4\,$eV, and $T=300\,$K. The dielectric spacer has thickness $d=1\,$$\mu$m and dielectric function $\epsilon_{\rm d}=2.1$, while for the gold substrate we use a Drude model fitted to the tabulated data in Ref.~[\citenum{johnson1972optical}]. (d) Reflection dissymmetry factor corresponding to the results of panel (c), calculated using Eq.~\eqref{eq:reflection_dissymmetry_factor}.}
\label{fig:fig2}
\end{center}
\end{figure}

We aim to describe the optical response of the system sketched in Fig.~\ref{fig:fig2}(a), which consists of a square array of graphene nanodisks, integrated with a dielectric spacer and a mirror into a Salisbury screen configuration. In our particular case, the dielectric spacer is made of a material with dielectric function $\epsilon_{\rm d}=2.1$, while the role of mirror is played by a gold substrate with dielectric function fitted from the data compiled in Ref.~[\citenum{johnson1972optical}]. When the period of the array is smaller than the wavelength of light, all diffraction orders except the zero order are evanescent. Then, following the approach of Ref.~[\citenum{thongrattanasiri2012complete}], we can neglect the contribution of the evanescent orders and calculate the response of the entire system using a transfer matrix method for stratified media in which the contribution of the array of graphene nanodisks is captured by a reflection coefficient (see the Appendix for details). 

Using this approach, we calculate the absorbance spectrum of the system for different values of the magnetic field. We consider an array with period $a=2D$, made of nanodisks with $D=60\,$nm, $E_{\rm F}=0.4\,$eV, and $T=300\,$K, along with a spacer of thickness $d=1\,$$\mu$m. As shown in Fig.~\ref{fig:fig2}(b), under zero magnetic field (black curve), the spectrum displays a single peak that reaches a maximum absorbance of approximately $99\%$, in agreement with Ref.~[\citenum{thongrattanasiri2012complete}]. The application of a magnetic field leads to the splitting of the spectrum into two peaks, corresponding to excitations with LCP (dashed curves) and RCP (solid curves) light. For all finite values of $B$ under consideration, both peaks in the spectrum reach near-perfect absorption, although their spectral overlap decreases as the magnetic field strength increases. These results are in excellent agreement with direct numerical solutions of Maxwell's equations as demonstrated in Fig.~\ref{fig:Salisbury_Screen_Comsol_comparison}. The reflectance of the array, which is plotted in Fig.~\ref{fig:fig2}(c), exhibits opposite behavior, approaching zero at the wavelengths of maximum absorption.

To quantify the strength of the chiral response of the system, we analyze in Fig.~\ref{fig:fig2}(d), the reflectance dissymmetry factor, defined as
\begin{equation} \label{eq:reflection_dissymmetry_factor}
g^{\rm ref} = 2\frac{\mathcal{R}_{+}-\mathcal{R}_{-}}{\mathcal{R}_{+}+\mathcal{R}_{-}}.
\end{equation}
Here, $\mathcal{R}_{+}$ and $\mathcal{R}_{-}$ correspond, respectively, to the reflectance for incoming LCP and RCP light. Examining the results displayed in Fig.~\ref{fig:fig2}(d), we observe that, for the three finite values of $B$ analyzed, the reflectance dissymmetry factor nearly reaches its theoretical maximum absolute value at the two peaks of the spectrum, indicating an almost perfectly chiral response. This is true even for $B=2.5\,$T, for which the absorbance spectrum shows a significant overlap between the LCP and RCP peaks. However, since both peaks achieve nearly perfect absorption, one polarization handedness is always effectively removed from the reflected light.

Kirchhoff's law \cite{greffet2018light} and its generalizations for magneto-optical systems \cite{khandekar2020new} establish a direct connection between the absorption properties of a structure and its capacity to emit thermal radiation. Therefore, the demonstrated ability to completely absorb light of only one handedness makes the system of Fig.~\ref{fig:fig2} a promising source of circularly polarized thermal radiation. To explore this possibility, we analyze the intensity of the thermal radiation emitted per unit frequency and solid angle in the direction perpendicular to the system. This quantity can be calculated as \cite{greffet2018light,khandekar2020new}
\begin{equation}
I_{\pm}(\omega)=\mathcal{E}_{\pm}(\omega)P_{\rm bb}(\omega), \nonumber
\end{equation}
where
\begin{equation}\label{eq:radiation_power_bb}
P_{\rm bb} (\omega) = \frac{\hbar \omega^3 }{8\pi^3 c^2}\frac{1}{{\rm e}^{\hbar\omega/k_{\rm B} T}-1},
\end{equation}
is the power emitted per unit frequency, solid angle, and area for a single polarization by a blackbody at temperature $T$, while $\mathcal{E}_{+}(\omega)$ and $\mathcal{E}_{-}(\omega)$ are the emissivities of the system for LCP  and RCP light, respectively. For a magneto-optical system such as the one considered here, the emissivity in the direction perpendicular to the system for one handedness is equal to the absorbance for the opposite handedness \cite{khandekar2020new}.

\begin{figure}
\begin{center}
\includegraphics[width=\linewidth]{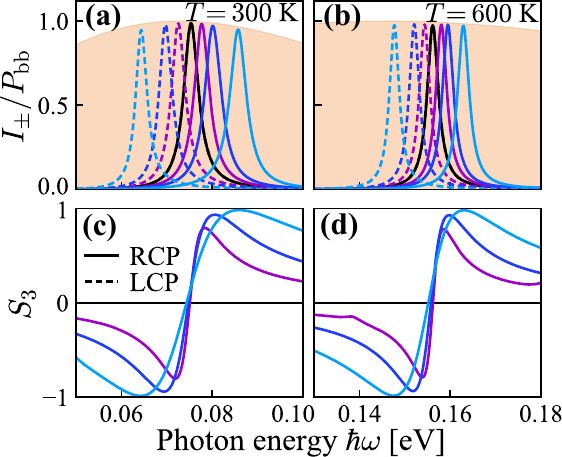}
\caption{{\bf Chiral thermal radiation.} (a,b) Intensity of the thermal radiation emitted per unit frequency and solid angle by the system of Fig.~\ref{fig:fig2}(a) in the direction perpendicular to it. In panel (a), we consider $a=1.8D$, $D=200\,$nm, $E_{\rm F}=0.3\,$eV, $d=2.625\,$$\mu$m, and $T=300\,$K, while in panel (b), $a=2D$, $D=60\,$nm, $E_{\rm F}=0.4\,$eV, $d=1\,$$\mu$m, and $T=600\,$K. The shaded area in the background signals $P_{\rm bb}(\omega)$ defined in Eq.~\eqref{eq:radiation_power_bb}. In both panels, we normalize the results to the maximum value of $P_{\rm bb}(\omega)$ for the same temperature. Dashed and solid curves correspond to LCP and RCP light, respectively, while the color represents the strength of $B$, as indicated by the legend in Fig.~\ref{fig:fig2}(d). (c,d) $S_3$ Stokes parameter for the thermal radiation of panels (a,b).}
\label{fig:fig3}
\end{center}
\end{figure}

Figures~\ref{fig:fig3}(a) and (b) display $I_\pm(\omega)$ for two different sets of parameters. Specifically, Fig.~\ref{fig:fig3}(a) corresponds to $a=1.8D$, $D=200\,$nm, $E_{\rm F}=0.3\,$eV, $d=2.625\,$$\mu$m, and $T=300\,$K, while for Fig.~\ref{fig:fig3}(b), $a=2D$, $D=60\,$nm, $E_{\rm F}=0.4\,$eV, $d=1\,$$\mu$m, and $T=600\,$K. In the absence of magnetic field (black curves), the thermal radiation spectrum displays a single peak, which splits into two for finite values of $B$.  As anticipated from the analysis of the absorbance, each of these peaks corresponds to the emission of light of a different handedness:  LCP (dashed curves) and RCP (solid curves). However, the LCP and RCP peaks are reversed with respect to the absorbance spectrum due to the relationship between emissivity and absorbance discussed above. Upon examining these peaks, we observe that in all cases considered in Figs.~\ref{fig:fig3}(a) and (b), the intensity nearly saturates the blackbody limit, which is signaled by the shaded area. It is important to remark that the dependence of the TMPs on the carrier concentration, magnetic field strength, and temperature, endows the system with an exceptional level of tunability. This flexibility ensures that it is always possible to design a Salisbury screen configuration such that the emission peak for the desired handedness matches the maximum of the blackbody spectrum. 

The analysis of the $S_3$ Stokes parameter, defined as
\begin{equation}
S_3 = \frac{\mathcal{E}_{-}(\omega)-\mathcal{E}_{+}(\omega)}{\mathcal{E}_{-}(\omega)+\mathcal{E}_{+}(\omega)}, \nonumber
\end{equation}
allows us to quantify the content of each polarization handedness in the thermal radiation emitted by the system. Figures~\ref{fig:fig3}(c) and (d)  
display $S_3$ for the parameters considered in Figs.~\ref{fig:fig3}(a) and (b), respectively. As expected, in the absence of magnetic field, $S_3=0$. However, for the smallest magnetic field strength considered, the absolute value of $S_3$ at the peaks approaches $0.8$. This value increases beyond $0.9$ for $B=5\,$T, and nearly reaches unity at $B=10\,$T, thus confirming the strong chiral nature of the thermal radiation emitted by the system.

In summary, we have investigated the optical response of doped graphene nanodisks at finite temperatures in the presence of a static magnetic field. We have found that the interplay between carrier concentration, magnetic field, and temperature gives rise to self-hybridized TMPs. These new modes originate from the hybridization of the magnetoplasmons supported by the nanodisks and  transitions between thermally populated LLs induced by the magnetic field. Remarkably, due to the quantized nature of the LLs, TMPs are enhanced rather than suppressed by temperature. Furthermore, TMPs exhibit near-perfect chiral responses, as demonstrated through our analysis of the extinction cross-section of individual nanodisks over a wide range of parameters. To obtain these results, we have developed a semianalytical formalism that accurately describes the eigenmodes and polarizability of graphene nanodisks for arbitrary values of carrier concentration, temperature, and magnetic field. We have demonstrated the excellent accuracy of this approach by benchmarking it against direct numerical solutions of Maxwell’s equations. 

As a direct application of our findings, we have shown that the TMPs supported by an array of graphene nanodisks in combination with a gold mirror are capable of perfectly absorbing light of a single handedness. Moreover, building on the known ability of graphene nanostructures to act as thermal emitters, we have demonstrated that the exact same configuration emits fully circularly polarized thermal radiation, whose handedness can be selected by tuning the characteristics of the TMPs. 
Our results highlight the potential of TMPs for enhancing and controlling chiral light absorption and emission, with the additional advantage of being tunable through three different mechanisms: carrier concentration, magnetic field, and temperature. 
Consequently, our work not only deepens the fundamental understanding of the optical properties of graphene nanostructures but also opens the door to their application in advanced photonic devices and technologies.

\begin{acknowledgements}
The authors thank Nuno M. R. Peres, Christos Tserkezis, and Eduardo J. C. Dias for insightful and enjoyable discussions.
J.~D.~C. is a Sapere Aude research leader supported by Independent Research Fund Denmark (Grant No. 0165-00051B).
The Center for Polariton-driven Light--Matter Interactions (POLIMA) is funded by the Danish National Research Foundation (Project No.~DNRF165).
J.~R.~D-R. and A.~M. acknowledge support from Grant No. PID2022-137569NB-C42 funded by MCIN/AEI/10.13039/501100011033. 
J.~R.~D-R. acknowledges support from a predoctoral fellowship from the MCIN/AEI assigned to Grant No. PID2019-109502GA-I00.
\end{acknowledgements}

\onecolumngrid
\appendix
\section{Appendix}\label{ap}

\section{Electrostatic Linear Response of Magneto-Optical Nanodisks}

We are interested in describing the optical response of a nanostructured 2D material located in the $z=0$ plane when excited by a monochromatic external field ${\bf E}^{\rm ext}{\rm e}^{-{\rm i}\omega t}+{\rm c.c.}$. Within the electrostatic approximation, the potential $\Phi$ in the 2D material is given by \cite{garcia2014graphene,yu2017analytical,muniz2020two}
\begin{equation} \label{eq:Phi_rho}
\Phi({\bf R},\omega) = \Phi^{\rm ext}({\bf R},\omega) + \frac{1}{\epsilon_{\rm eff}}\int {\rm d}^2{\bf R}'\frac{\rho({\bf R}',\omega)}{\left|{\bf R}-{\bf R}'\right|} ,
\end{equation}
where $\Phi^{\rm ext}$ is the potential associated with the external field and $\rho({\bf R},\omega)$ is the 2D charge density given at position ${\bf R}=(x,y)$. Furthermore, $\epsilon_{\rm eff}=(\epsilon_{\rm a}+\epsilon_{\rm b})/2$ is the effective dielectric function defined in terms of the dielectric function of the medium above, $\epsilon_{\rm a}$, and below, $\epsilon_{\rm b}$, the 2D material. Invoking the continuity equation $\nabla_{\bf R}\cdot{\bf J}=-\partial \rho / \partial t$ and Ohm's law ${\bf J}=\sigma\cdot{\bf E}$, the charge density can be expressed as
\begin{equation} \label{eq:rho_Phi}
\rho({\bf R},\omega) = \frac{{\rm i}}{\omega}\nabla_{\bf R}\cdot\left[\sigma({\bf R},\omega)\cdot\nabla_{\bf R}\Phi({\bf R},\omega)\right] ,
\end{equation}
where we introduce the 2D conductivity tensor $\sigma({\bf R},\omega)$ characterizing the optical response of the 2D material. In the local limit, we express $\sigma({\bf R},\omega)=\Theta({\bf R})\sigma(\omega)$, where $\Theta({\bf R})$ is a geometrical factor that equals unity for coordinates ${\bf R}$ within the 2D structure and vanishes otherwise, while $\sigma(\omega)$ represents the intrinsic conductivity tensor of the infinitely extended  2D material.

The case of a magneto-optical 2D nanodisk of diameter $D$ is conveniently treated in cylindrical coordinates defined by the radial distance $r$ and azimuthal angle $\varphi$. Exploiting the cylindrical symmetry of the structure, we decompose the potential and charge density as $\Phi({\bf R})=\sum_{l=-\infty}^\infty\phi_l(r){\rm e}^{{\rm i} l\varphi}$ and $\rho({\bf R})=\sum_{l=-\infty}^\infty\rho_l(r){\rm e}^{{\rm i} l\varphi}$, respectively, while the Coulomb interaction can be expanded as \cite{fetter1986magnetoplasmons, christensen2014classical,muniz2020two}
\begin{equation}
\frac{1}{\left|{\bf R}-{\bf R}'\right|} = \sum_{l=-\infty}^\infty\int_0^\infty {\rm d}q J_{\left|l\right|}(qr)J_{\left|l\right|}(qr'){\rm e}^{{\rm i} l(\varphi-\varphi')}. \nonumber
\end{equation}
Here, $J_l$ represents the cylindrical Bessel function of order $l$. Inserting these expressions into Eq.~\eqref{eq:Phi_rho}, we get 
\begin{equation}\label{eq:phil_philext_rhol}
\phi_l(r) = \phi_l^{\rm ext}(r) + \frac{2\pi}{\epsilon_{\rm eff}}\int_0^\infty {\rm d}r'r' \int_0^\infty {\rm d}q J_{\left|l\right|}(qr)J_{\left|l\right|}(qr') \rho_l(r') ,
\end{equation}
with
\begin{equation}
\phi_l^{\rm ext}(r) = \frac{1}{2\pi}\int_0^{2\pi} {\rm d}\varphi \Phi^{\rm ext}({\bf R}){\rm e}^{-{\rm i} l\varphi}.\nonumber
\end{equation}
Meanwhile, the charge density of Eq.~\eqref{eq:rho_Phi}, expressed in cylindrical coordinates, becomes
\begin{equation} \label{eq:rho_l}
\rho_l(r) = \frac{{\rm i}}{\omega}\left\{\Theta(r)\sigma_{\parallel}\left[\frac{1}{r}\frac{\partial}{\partial r}\left(r\frac{\partial}{\partial r}\right) - \frac{l^2}{r^2}\right] + \delta(r-D/2)\left(\sigma_{\parallel}\frac{\partial}{\partial r} + \sigma_{\perp}\frac{{\rm i} l}{r}\right)\right\}\phi_l(r),
\end{equation}
where $\Theta(r)$ is the radial version of $\Theta({\bf R})$ and the components of the conductivity tensor of the 2D magneto-optical material satisfy $\sigma_{rr}=\sigma_{\varphi\varphi} \equiv \sigma_{\parallel}$ and $\sigma_{r\varphi}=-\sigma_{\varphi r} \equiv \sigma_{\perp}$. This expression indicates that the Hall conductivity, $\sigma_{\perp}$, only contributes at the edge of the charge distribution, with the sign of $l$ determining the direction of the current flow.

Following the prescription of Ref.~\cite{fetter1986magnetoplasmons}, we define $x = 2r/D$ and introduce the Green function $G_l(x,x')$ that satisfies
\begin{equation}
\left[\frac{1}{x}\frac{\partial}{\partial x}\left(x\frac{\partial}{\partial x}\right)-\frac{|l|^2}{x^2}\right]G_l(x,x') = -\frac{1}{x}\delta(x-x'), \nonumber
\end{equation}
on the interval $[0,1]$, together with the two boundary conditions that $G(0,x')$ is bounded and that
\begin{equation} \label{eq:G_BC}
\left[\left(\sigma_{\parallel}\frac{\partial}{\partial x} + \sigma_{\perp}\frac{{\rm i} l}{x}\right)G_l(x,x')\right]_{x\to1^-} = 0.
\end{equation}
This allows us to recast Eq.~\eqref{eq:rho_l} in integral form as
\begin{equation}\label{eq:phil_Gl_rhol}
\phi_l(x) = \frac{{\rm i}\omega D^2}{4\sigma_{\parallel}}\int_0^1 {\rm d}x'x'G_l(x,x')\rho_l(x')
\end{equation}

For $l\neq0$, the explicit form of $G_l(x,x')$ is determined from the ansatz
\begin{equation}
G_l(x,x') = \frac{1}{2|l|}\left[A_l(xx')^{|l|} + B_l\left(\frac{x_<}{x_>}\right)^{|l|}\right],\nonumber
\end{equation}
where $x_<={\rm min}(x,x')$ and $x_>={\rm max}(x,x')$, while the coefficients $A_l$ and $B_l$ are obtained from the boundary condition in Eq.~\eqref{eq:G_BC} as $B_l=1$ and $A_l = [\sigma_{\parallel}-{\rm i}\,{\rm sgn}(l)\sigma_{\perp}]/[\sigma_{\parallel}+{\rm i}\,{\rm sgn}(l)\sigma_{\perp}]$, so \cite{fetter1986magnetoplasmons}
\begin{equation}
G_l(x,x') = \frac{1}{2|l|}\left[\frac{\sigma_{\parallel}-{\rm i}\,{\rm sgn}(l)\sigma_{\perp}}{\sigma_{\parallel}+{\rm i}\,{\rm sgn}(l)\sigma_{\perp}}(xx')^{|l|} + \left(\frac{x_<}{x_>}\right)^{|l|}\right]. \nonumber
\end{equation}

Combining Eqs.~\eqref{eq:phil_philext_rhol} and \eqref{eq:phil_Gl_rhol}, we obtain an integral equation for the charge density
\begin{equation} \label{eq:Full_integral_eq}
\frac{D}{2}\int_0^1 {\rm d}x' x' \left[\frac{2\pi}{\epsilon_{\rm eff}}\int_0^\infty {\rm d}p J_{|l|}(xp)J_{|l|}(x'p) - \frac{{\rm i}\omega D}{2\sigma_{\parallel}}G_l(x,x')\right]\rho_l(x') = -\phi^{\rm ext}_l(x),
\end{equation}
which can be transformed to a matrix equation by expanding the charge density as \cite{fetter1986magnetoplasmons}
\begin{equation} \label{eq:rho_l_Jacobi}
\rho_l(x) = \sum_{j=0}^\infty a_j^{(l)} x^{|l|}P_j^{(|l|,0)}(1-2x^2).
\end{equation}
Here, $a_j^{(l)}$ are expansion coefficients and $P_j^{(|l|,0)}(1-2x^2)$ are Jacobi polynomials that satisfy the orthonormality condition
\begin{equation}
\int_0^1 {\rm d}x x^{2|l|+1}P_j^{(|l|,0)}(1-2x^2)P_{j'}^{(|l|,0)}(1-2x^2) = \frac{\delta_{jj'}}{2(|l|+2j+1)}, \nonumber
\end{equation}
as well as the integral relation
\begin{equation}
\int_0^1 {\rm d}xx^{|l|+1}J_{|l|}(xp)P_j^{(|l|,0)}(1-2x^2) = \frac{1}{p}J_{|l|+2j+1}(p). \nonumber
\end{equation}
Then, making use of the conditions \cite{gradshteyn2014table}
\begin{equation}
 \int_0^\infty \frac{{\rm d}p}{p}J_{|l|+2j+1}(p)J_{|l|+2j'+1}(p) = \frac{\delta_{jj'}}{2(|l|+2j+1)},\nonumber
\end{equation}
and
\begin{equation}
 \int_0^\infty \frac{{\rm d}p}{p^2}J_{|l|+2j+1}(p)J_{|l|+2j'+1}(p) = \frac{(-1)^{j-j'+1}}{\pi[4(j-j')^2-1](|l|+j+j'+\tfrac{1}{2})(|l|+j+j'+\tfrac{3}{2})} ,\nonumber
\end{equation}
the recursion relation
\begin{equation}
\frac{2m}{p} J_m(p)=J_{m-1}(p)+J_{m+1}(p), \nonumber
\end{equation}
and the identity
\begin{equation}
\frac{1}{2|l|}\left(\frac{x_<}{x_>}\right)^{|l|} = \int_0^\infty\frac{{\rm d}p}{p}J_{|l|}(px)J_{|l|}(px') \nonumber
\end{equation}
for the cylindrical Bessel functions, we can multiply Eq.~\eqref{eq:Full_integral_eq} by $x^{|l|+1}P_{j'}^{(|l|,0)}(1-2x^2)$ and integrate over $x$ to obtain the matrix equation 
\begin{equation} \label{eq:General_matrixEOM_disk}
    \left(\chi^{(l)}\mathcal{A}^{(l)} + \eta\mathcal{B}^{(l)} - \mathcal{C}^{(l)}\right)\cdot {\bf a}^{(l)} = {\bf b}^{(l)}.
\end{equation}
In this expression, $\eta = {\rm i}\epsilon_{\rm eff} \omega D/ 4\pi \sigma_{\parallel}$ and 
\begin{equation}
\chi^{(l)} = \eta \frac{\sigma_{\parallel}-{\rm i}\,{\rm sgn}(l)\sigma_{\perp}}{\sigma_{\parallel}+{\rm i}\,{\rm sgn}(l)\sigma_{\perp}} \nonumber
\end{equation}
include all the material-dependent parameters of the problem. The elements of the three matrices are defined as 
\begin{align}
\mathcal{A}^{(l)}_{jj'} &= \frac{\delta_{0j}\delta_{0j'}}{8|l|(|l|+1)^2} ,  \nonumber \\
\mathcal{B}^{(l)}_{jj'} &= \frac{1}{8(|l|+2j+1)(|l|+2j'+1)}\left(\frac{\delta_{jj'}+\delta_{j, j'+1}}{|l|+2j} + \frac{\delta_{jj'}+\delta_{j+1,j'}}{|l|+2j+2}\right) , \nonumber  \\
\mathcal{C}^{(l)}_{jj'} &= \frac{(-1)^{j-j'+1}}{\pi[4(j-j')^2-1](|l|+j+j'+\tfrac{1}{2})(|l|+j+j'+\tfrac{3}{2})}, \nonumber
\end{align}
while the vector ${\bf a}^{(l)}$ contains the coefficients $a_j^{(l)}$ introduced in Eq.~\eqref{eq:rho_l_Jacobi} and 
\begin{equation}
b_j^{(l)} = \frac{\epsilon_{\rm eff}}{\pi D}\int_0^1 {\rm d}xx^{|l|+1}P_j^{(|l|,0)}(1-2x^2)\phi^{\rm ext}_l(x) \nonumber
\end{equation}
encodes the external potential. Therefore, for a given $\phi^{\rm ext}_l(x)$, we can solve Eq.~\eqref{eq:General_matrixEOM_disk} to obtain the charge induced in the nanodisk. In the absence of an external potential, Eq.~\eqref{eq:General_matrixEOM_disk} reduces to an eigenvalue problem that can be solved by computing a determinant $\left|\mathcal{M}^{(l)}\right|=0$ with
\begin{equation} \label{eq:M_l}
\mathcal{M}^{(l)} =  \mathcal{C}^{(l)} -\chi^{(l)}\mathcal{A}^{(l)} - \eta\mathcal{B}^{(l)} . \nonumber
\end{equation}
While all the matrices defined above are infinite, we find well-converged results by truncating them after $j,j'\sim 300$ for the systems considered in this work.

Although not directly relevant to our study,  it can be shown that the Green function for $l=0$, which is given by $G_0(x,x')=-\ln(x_{>})$, results in the same matrix equation, except that $\mathcal{A}_{00}^{(0)}=0$, and the other matrices omit the terms corresponding to $j=j'=0$, starting instead at $j,j'=1$ \cite{fetter1986magnetoplasmons,muniz2020two}.

The polarizability of the nanodisk is computed by decomposing the potential associated with the external electric field ${\bf E}^{\rm ext}$ into left- (LCP) and right-handed (RCP) circularly polarized components as
\begin{equation}
 \Phi^{\rm ext}({\bf R}) = -\frac{1}{2}E^{\rm ext}r\left({\rm e}^{{\rm i}\varphi}+{\rm e}^{-{\rm i}\varphi}\right), \nonumber
\end{equation}
where we assume that the external field is polarized along $\hat{\bf x}$. The corresponding solution of Eq.~\eqref{eq:General_matrixEOM_disk} for this potential is given by
\begin{equation}
{\bf a}^{(l)} = \frac{\epsilon_{\rm eff}}{16\pi}\left(\delta_{l,1}+\delta_{l,-1}\right)[\mathcal{M}^{(l)}]^{-1}\cdot\hat{\bf e}_0 E^{\rm ext}, \nonumber
\end{equation}
where $\hat{\bf e}_0$ is a vector with 1 in the zeroth entry and 0 in all other entries. Once the coefficients $a_j^{(l)}$ are known, the induced charge density is computed as
\begin{equation}
\rho(x,\varphi) = \sum_{l=-\infty}^{\infty}\sum_{j=0}^{\infty} a_j^{(l)} x^{|l|} P_j^{(|l|,0)} (1-2x^2) {\rm e}^{{\rm i} l\varphi}, \nonumber
\end{equation}
which, in turn, allows us to obtain the induced dipole as  ${\bf p} = \int {\rm d}^2{\bf R}{\bf R}\rho({\bf R})$. 
Repeating this derivation for an external field polarized along $\hat{\bf y}$, the polarizability tensor of the nanodisk, defined from ${\bf p} = \alpha_0\cdot{\bf E}^{\rm ext}$, can be written as 
\begin{equation}
\alpha_0 = \frac{\epsilon_{\rm eff}}{64}\left(\frac{D}{2}\right)^3\left[M^{(+1)}\begin{pmatrix}
        1 & -{\rm i} \\ {\rm i} & 1
\end{pmatrix} + M^{(-1)}\begin{pmatrix}
        1 & {\rm i} \\ -{\rm i} & 1
\end{pmatrix}\right], \nonumber
\end{equation}
with $M^{(l)}\equiv\hat{\bf e}_0\cdot\left[\mathcal{M}^{(l)}\right]^{-1}\cdot\hat{\bf e}_0$. The two terms of this polarizability represent the coupling to LCP and RCP light. For a graphene nanodisk subjected to a perpendicular static magnetic field, this expression predicts a spectral splitting of the magnetoplasmon peak in the extinction cross-section, in alignment with previous theoretical \cite{wang2012edge,wang2011edge} and  experimental \cite{yan2012infrared, poumirol2017electrically} results. As expected, the peaks coalesce into a single peak when the magnetic field is zero.

\section{Electrodynamic Corrections to the Polarizability}

Although the nanodisk polarizability discussed above is derived in the electrostatic regime, we can include electrodynamic corrections that account for retardation effects by following the prescription of Ref.~\cite{deop2022optical}. Specifically, we begin by introducing a depolarization field ${\bf E}^{\rm d}$, such that the induced dipole satisfies
\begin{equation} \label{eq:p_dep}
 {\bf p} = \alpha_0 \cdot \left( {\bf E}^{\rm ext} + {\bf E}^{\rm d} \right). 
\end{equation}
Then, assuming that the induced dipole moment is uniformly distributed across the nanodisk, the depolarization field is given by \cite{moroz2009depolarization}
\begin{equation} \label{eq:e_dep}
{\bf E}^{\rm d} =  \left( \frac{3k^2}{D} + {\rm i} \frac{2k^3}{3} \sqrt{\epsilon} \right) {\bf p}, 
\end{equation}
with $k=\omega/c$. This expression is valid when the nanodisk is surrounded by a homogeneous medium with dielectric function $\epsilon$. However, as we show below, it also serves as a good approximation for a nanodisk located at the interface between two media when $\epsilon$ is substituted by $\epsilon_{\rm eff}$. Then, substituting Eq.~\eqref{eq:e_dep} into Eq.~\eqref{eq:p_dep}, the corrected polarizability reads
\begin{equation}
\alpha=\left[\alpha_0^{-1} - k^2\left(\frac{3}{D} + \frac{2}{3}{\rm i} k \sqrt{\epsilon_{\rm eff}}\right)\mathbbm{1}_{2\times2}\right]^{-1}, \nonumber
\end{equation}
where $\mathbbm{1}_{2\times2}$ is the identity matrix of dimension $2$.

\section{Magneto-Optical Conductivity of Graphene}

In the presence of a normally-oriented static magnetic field $B$, the charge carriers in graphene occupy quantized Landau levels (LLs) with energies
\begin{equation}
E_n = \text{sgn}(n) \frac{\hbar v_{\rm F}}{L_B} \sqrt{2|n|}, \nonumber
\end{equation}
where $n\in\{0,\,\pm 1,\,\pm 2, ... \}$ indexes the level, $v_{\rm F}$ is the Fermi velocity of this material, and $L_B=\sqrt{\hbar c/eB}$ is the magnetic length \cite{ferreira2011faraday,sounas2011electromagnetic,sounas2012gyrotropy,gonccalves2016introduction}. Following the method of Ref.~\cite{ferreira2011faraday}, the longitudinal component of the optical conductivity is computed as a sum over LL transitions according to 
\begin{equation} \label{eq:sigma_xx_B}
\sigma_{xx}(\omega) = \sigma_{yy}(\omega) = \frac{{\rm i} e^2}{2\pi\hbar} \sum_{n \neq m =-N_c}^{N_c} \frac{\Lambda_{nm}^{xx}}{E_{nm}} \frac{f_m-f_n}{\hbar \omega + E_{nm} + {\rm i} \hbar/\tau} ,
\end{equation}
where $f_n$ is the Fermi-Dirac occupation factor of the $n^{\rm th}$ LL,
\begin{equation}
\Lambda_{nm}^{xx} = \frac{\hbar^2 v_F^2}{L_B^2} (1 + \delta_{m,0}+\delta_{n,0})\delta_{|m|-|n|,\pm 1}, \nonumber
\end{equation}
$E_{nm} = E_n-E_m$, and the summation over LLs is truncated at an index $N_c = \text{int}[E_{c}^2/E_1^2]$, where $E_{c}$  is a cutoff energy. Throughout this work we choose $E_{c}=2.7\,$eV corresponding to the nearest-neighbor tight-binding hopping energy of graphene \cite{ferreira2011faraday}. For finite temperature, $f_n$, is found from $f_n=f(E_n) = [{\rm e}^{(E_n-\mu)/k_{\rm B} T}+1]^{-1}$ for a chemical potential $\mu$ and temperature $T$, while for zero temperature the LLs are filled up from the lowest LL and up until all available charges are used. Similarly, the Hall conductivity is given by
\begin{equation} \label{eq:sigma_xy_B}
\sigma_{xy}(\omega) = -\sigma_{yx}(\omega) = \frac{{\rm i} e^2}{2\pi\hbar} \sum_{n \neq m =-N_c}^{N_c} \frac{\Lambda_{nm}^{xy}}{E_{nm}} \frac{f_m-f_n}{\hbar \omega + E_{nm} +  {\rm i}  \hbar/\tau},
\end{equation}
with
\begin{equation}
\Lambda_{nm}^{xy} = {\rm i} \Lambda_{nm}^{xx} (\delta_{|m|,|n|-1}- \delta_{|m|-1,|n|}). \nonumber
\end{equation}

In the absence of a magnetic field, the conductivity becomes isotropic, i.e., $\sigma_{xy}=\sigma_{yx}=0$ and $\sigma_{xx} = \sigma_{yy}$, and, within the local RPA, is given by \cite{yu2017ultrafast}
\begin{equation}
\sigma_{xx}(\omega)=\sigma_{yy}(\omega) = \frac{{\rm i} e^2}{\pi \hbar^2} \frac{1}{\omega+{\rm i}\tau^{-1}} \left\{ \mu_{\rm D} - \int_0^{\infty}{\rm d}E \frac{f(E)-f(-E)}{1-4E^2 /[\hbar^2(\omega+ {\rm i}/\tau)^2]} \right\}. \nonumber
\end{equation}
Here, the first term accounts for the intraband response of charge carriers with effective Drude weight
\begin{equation}
\mu_{\rm D} = \mu + 2k_{\rm B} T \log\left( 1 + {\rm e}^{-\mu/k_{\rm B}T} \right),\nonumber
\end{equation}
while the second term describes the response of the interband transitions.

\section{Dependence of the Chemical Potential on the Temperature and the Magnetic Field}

\begin{figure} 
\begin{center}
\includegraphics[width=\linewidth]{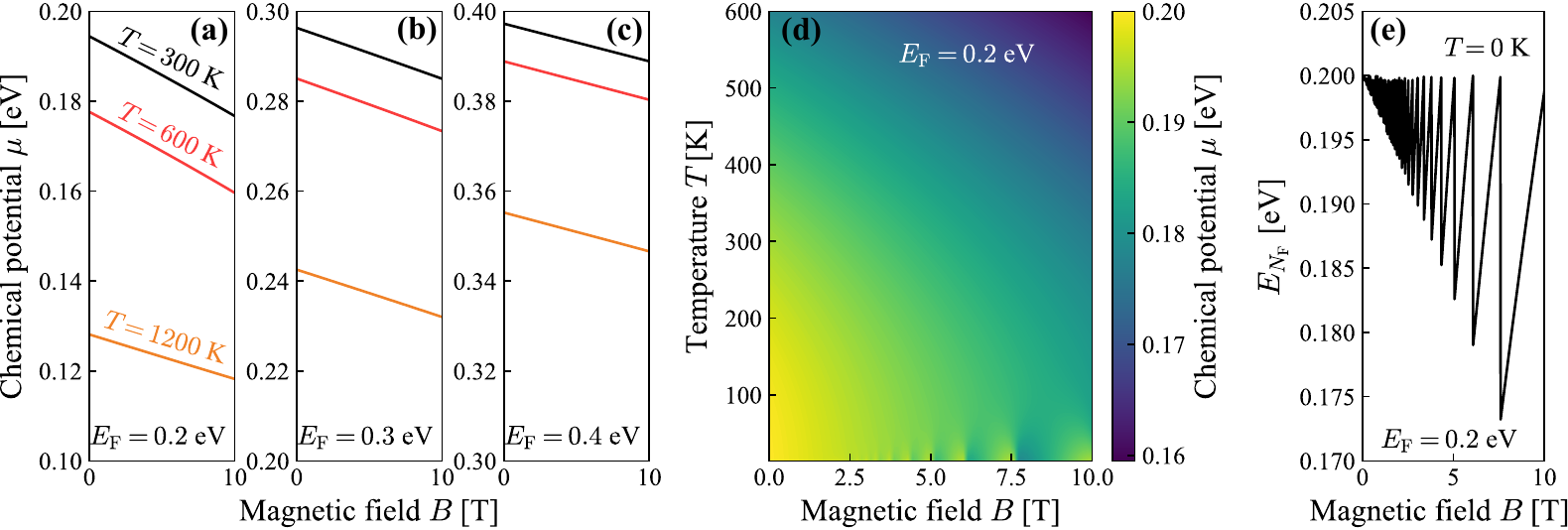}
\caption{{\bf Chemical potential of graphene as a function of magnetic field and temperature.} We compute $\mu$ as a function of $B$ for three temperatures $T=300\,$K, $T=600\,$K and $T=1200\,$K, as indicated by the legend, and three Fermi energies $E_{\rm F}=0.2\,$eV  (a), $E_{\rm F}=0.3\,$eV  (b), and $E_{\rm F}=0.4\,$eV (c). (d) Contour plot showing the chemical potential as a function of temperature and magnetic field for $E_{\rm F}=0.2\,$eV. (e) Energy of the highest occupied LL  at zero temperature and $E_{\rm F}=0.2\,$eV as computed from Eq.~\eqref{eq:Chemical_potential_BField_num_find}.}
\label{fig:Chemical_pot}
\end{center}
\end{figure}

In the absence of a magnetic field, the linear electronic dispersion relation of graphene $E=\hbar v_{\rm F}|{\bf k}|$ leads to the charge carrier density \cite{yu2017ultrafast}
\begin{equation} \label{eq:n}
n = \frac{g_{\rm v} g_{\rm s}}{2\pi \hbar^2 v_{\rm F}^2} \int_0^{\infty} {\rm d}E E \left[f(E) + f(-E)-1 \right],
\end{equation}
where $g_{\rm s}=2$ and $g_{\rm v}=2$ account for spin and valley degeneracy, respectively, and the Fermi-Dirac distribution $f(E)$ also depends on chemical potential $\mu$ and temperature $T$. From the zero-temperature definition of the Fermi energy $E_{\rm F}=\hbar v_{\rm F}\sqrt{\pi n}$, Eq.~\eqref{eq:n} establishes a relation between $E_{\rm F}$, $\mu$, and $T$ according to \cite{yu2017ultrafast,cox2019single}
\begin{equation}
\left( \frac{E_{\rm F}}{k_{\rm B} T} \right)^2 = 2 \int_0^{\infty} {\rm d}x x\left( \frac{1}{{\rm e}^{x-\mu/k_{\rm B}T}+1}-\frac{1}{{\rm e}^{x+\mu/k_{\rm B}T}+1} \right) = 2 \left[ {\rm Li}_2\left( - {\rm e}^{-\mu/k_{\rm B}T} \right)-{\rm Li}_2\left( - {\rm e}^{\mu/k_{\rm B}T} \right) \right], \nonumber
\end{equation}
where ${\rm Li}_s(x)$ denotes the polylogarithm function of order $s$. 

In the presence of a perpendicular magnetic field $B$, the linear electron dispersion relation of graphene splits into LLs with degeneracy $\mathcal{S}/2\pi L_B^2$, where $\mathcal{S}$ is the area of the graphene sheet \cite{ferreira2011faraday}. In this situation, the electron density is
\begin{equation}
n = \frac{g_{\rm v}g_{\rm s}}{2\pi}\frac{eB}{c \hbar} \sum_{n=0}^{\infty}\left[ f_n-f_{-n}+1 \right]. \nonumber
\end{equation}
Introducing the same cutoff $N_{\rm c}$ as in Eqs.~\eqref{eq:sigma_xx_B} and \eqref{eq:sigma_xy_B}, and enforcing the conservation of the charge density at zero magnetic field, we obtain
\begin{equation}\label{eq:Chemical_potential_BField_num_find}
\frac{E_{\rm F}^2}{\hbar v_{\rm F}^2} = \frac{2eB}{c}\sum_{n=0}^{N_{\rm c}}\left[ f_n-f_{-n}+1 \right].
\end{equation}
which, for finite temperatures, can be written as
\begin{equation} \label{eq:Chemical_potential_BField_finiteT_num_find}
\frac{E_{\rm F}^2}{\hbar v_{\rm F}^2} = \frac{2eB}{c}\sum_{n=0}^{N_{\rm c}}  \left[ \frac{1}{{\rm e}^{(E_n-\mu)/k_{\rm B}T}+1}-\frac{1}{{\rm e}^{(E_n+\mu)/k_{\rm B}T}+1} \right].
\end{equation}
Therefore, given a Fermi energy and a temperature, the chemical potential is determined numerically from the expressions above. In Figs.~\ref{fig:Chemical_pot}(a)-(c), we analyze the dependence of the chemical potential on the magnetic field for different temperatures and Fermi levels considered in the main text. As a general trend, $\mu$ decreases as the magnetic field increases. For low temperatures, we observe in Fig.~\ref{fig:Chemical_pot}(d) that $\mu$ oscillates with $B$.  These oscillations are associated with the strong variations of the highest occupied LL energy, which are illustrated in Fig.~\ref{fig:Chemical_pot}(e) for $E_{\rm F}=0.2\,$eV and $T=0\,$K, and result from the interplay of carrier conservation and the LL degeneracy.

\section{Temperature Dependence of the Electron Damping Rate}

Although we assume a constant value for the electron damping rate $\tau^{-1}$ in the main text for simplicity, it is known that $\tau^{-1}$ increases as the temperature of the system rises, primarily due to the enhancement of electron-phonon scattering \cite{sohier2014phonon}. To model this phenomenon, we follow the theoretical approach developed in Ref~\cite{sohier2014phonon}, where the authors provide analytical expressions with parameters obtained from density functional theory (DFT) calculations to describe the contributions of the different electron-phonon scattering channels to $\tau^{-1}$. In the following, we reproduce these expressions, incorporating the distinction between the electron temperature $T_e$ and the lattice temperature $T_l$. 
The approach starts by considering the energy-dependent electron damping rate $\tau(E)$ entering the Boltzmann transport equation, which can be approximated by summing over different decay channels using Matthiessen's rule as
\begin{equation}
    \frac{1}{\tau(E)} = \sum_{\nu} \frac{1}{\tau_{\nu}(E)}. \nonumber
\end{equation}
Then, within the high-temperature regime, which is defined as $T \gtrsim 270\,$K and corresponds to the limit where the elastic approximation for acoustic phonons is valid, it is possible to obtain the following analytical expressions \cite{sohier2014phonon}. 
The contribution from the acoustic phonons is given by
\begin{equation}
    \frac{1}{\tau_{\rm A} (E)} = \frac{2\beta_{\rm A}^2 k_{\rm B} T_l}{\mu_{\rm s} \hbar v_{\rm A}^2} \frac{E}{(\hbar v_{\rm F})^2}, \nonumber
\end{equation}
where $\beta_{\rm A}$ is the gauge field as computed from DFT in Table IV of Ref.~\cite{sohier2014phonon}, $\mu_{\rm s}$ is the surface mass density of graphene, and $v_{\rm A}$ is the effective sound velocity. For the contribution of the optical LO and TO phonons, with the assumption of a constant optical phonon energy $\hbar\omega_{\rm O} = 0.2\,$eV, we have
\begin{equation}
    \frac{1}{\tau_{\rm O}(E)} = \frac{\beta_{\rm O}^2}{\mu_{\rm s} \omega_{\rm O} (\hbar v_{\rm F})^2} \left[ n_{\rm O} |E + \hbar \omega_{\rm O}| \frac{1-f(E + \hbar \omega_{\rm O})}{1-f(E)} +(n_{\rm O}+1)|E -\hbar \omega_{\rm O}| \frac{1-f(E-\hbar\omega_{\rm O})}{1-f(E)}\right], \nonumber
\end{equation}
where $\beta_{\rm O}$ is the corresponding gauge field, and $n_{\rm O}= \left[{\rm e}^{\hbar \omega_{\rm O}/k_{\rm B}T_l}-1\right]^{-1}$. Similarly, for the optical $A'_1$ phonons, assuming a constant energy $\hbar\omega_{\rm A'_1}=0.15\,$eV, we can write
\begin{align}
    \frac{1}{\tau_{\rm A'_1}(E)} ={}& \frac{\beta_{\rm K}^2  }{\mu_{\rm s} \omega_{\rm A'_1} (\hbar v_{\rm F})^2} \left\{ \frac{3}{2} n_{\rm A'_1} |E + \hbar \omega_{\rm A'_1}| \frac{1-f(E + \hbar \omega_{\rm A'_1})}{1-f(E)} \right. \nonumber\\
    &\left.+(n_{\rm A'_1}+1)\left[|E -\hbar \omega_{\rm A'_1}| + \frac{1}{2}(E - \hbar\omega_{\rm A'_1}) \right] \frac{1-f(E-\hbar\omega_{\rm A'_1})}{1-f(E)}\right\}, \nonumber
\end{align}
where, as before, $\beta_{\rm K}$ is a parameter, and $n_{\rm A'_1}= \left[{\rm e}^{\hbar \omega_{\rm A'_1}/k_{\rm B}T_l}-1\right]^{-1}$. We note that the values of $\beta_{\rm A}$, $\beta_{\rm O}$, and $\beta_{\rm K}$, as well as $\mu_{\rm s}$ and $v_{\rm A}$, are provided in Table IV of Ref.~\cite{sohier2014phonon}. 

The resulting contribution to the electron damping rate due to the different electron-phonon scattering channels is \cite{yuan2020room,dias2020thermal}
\begin{equation}
    \frac{1}{\tau_{\rm p}(E)} = \frac{1}{\tau_{\rm A} (E)} + \frac{1}{\tau_{\rm O}(E)} + \frac{1}{\tau_{\rm A'_1}(E)}.  \nonumber
\end{equation}
Using Matthiessen's rule, we can add this contribution to the constant damping rate $\tau_{0}^{-1}= e v_{\rm F}^2 / \mu_{\rm DC} E_{\rm F}$ that we use in the main text, which here we denote as $\tau^{-1}_0$ to avoid confusion. By taking this conservative approach, we get
\begin{equation}
    \frac{1}{\tau(E)}  = \frac{1}{\tau_{0}}+  \frac{1}{\tau_{\rm p}(E)}. \nonumber
\end{equation}
Then, following Ref.~\cite{dias2020thermal}, the total electron damping rate is given by
\begin{equation}
    \frac{1}{\tau} = \frac{\int {\rm d}E D(E) \partial_E f(E)}{\int {\rm d}E D(E)  \tau(E) \partial_E f(E)}, \label{eq:weighted_thermalaverage_scattering}
\end{equation}
where the electronic density of states is $D(E)=2|E|/\pi\hbar^2 v_{\rm F}^2$ and the Fermi-Dirac distributions are evaluated using the electron temperature $T_e$. 

When a static magnetic field is present, we can generalize the weighted average of Eq.~\eqref{eq:weighted_thermalaverage_scattering} 
\begin{equation}
    \frac{1}{\tau} = \frac{\sum_{n=0}^{\infty} 
 \left[\partial_E f(E)\right]_{E=E_n}}{\sum_{n=0}^{\infty} \tau(E_n)
 \left[\partial_E f(E)\right]_{E=E_n}}.  
\label{eq:weighted_thermalaverage_scattering_Bfield}
\end{equation}

\begin{figure} 
\begin{center}
\includegraphics[width=0.659\linewidth]{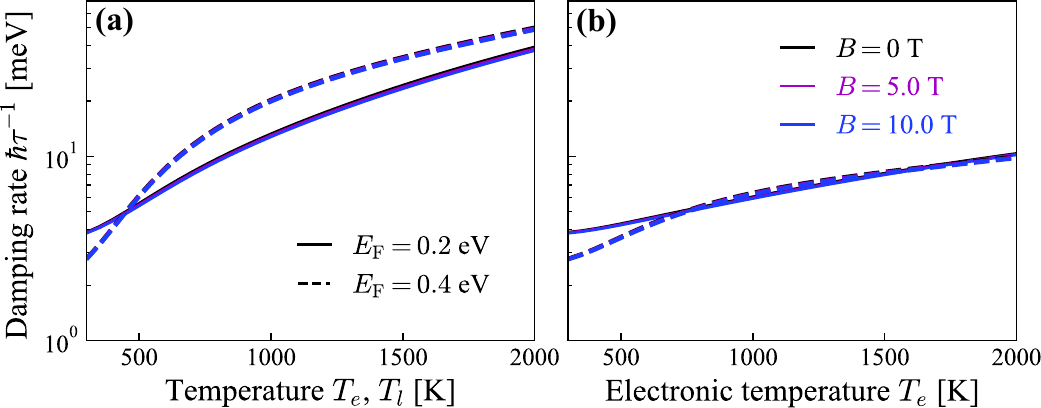}
\caption{ {\bf Temperature-dependent electron damping rate.} (a) Electron damping rate calculated from Eqs.~\eqref{eq:weighted_thermalaverage_scattering} and \eqref{eq:weighted_thermalaverage_scattering_Bfield} as a function of the electronic temperature $T_e$, assuming that the lattice temperature satisfies $T_l=T_e$. Solid and dashed curves correspond to ${E_{\rm F}}=0.2\,$eV and ${E_{\rm F}}=0.4\,$eV, respectively, while the different colors indicate the magnetic field strength, as shown in the legend. (b) Same as (a) but assuming that the lattice temperature remains constant at $T_l=300\,$K.} \label{fig:temp_dependent_scattering_rate}
\end{center}
\end{figure}

\begin{figure} 
\begin{center}
\includegraphics[width=0.654\linewidth]{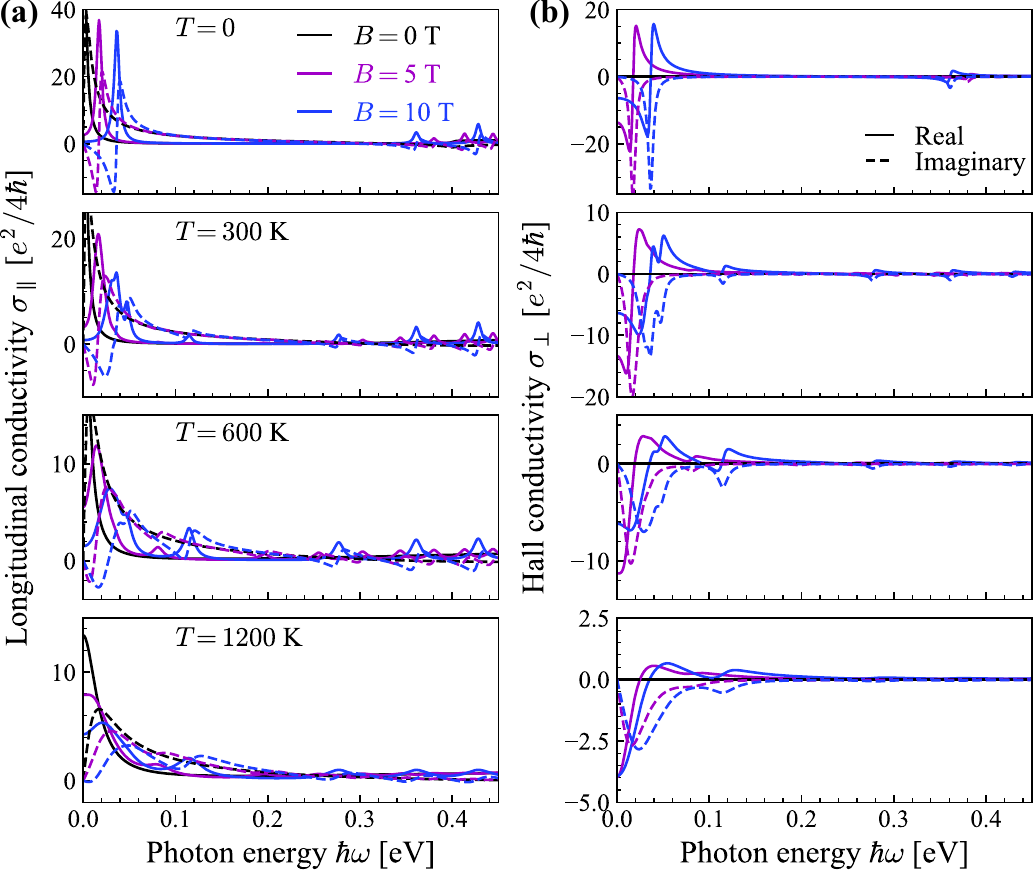}
\caption{{\bf Effect of temperature and magnetic field on graphene conductivity.} (a) Longitudinal component of the conductivity $\sigma_{xx}$, as computed from Eq.~\eqref{eq:sigma_xx_B} for different values of magnetic field and temperatures, as indicated by the labels. Solid and dashed curves correspond to the real and imaginary part of the conductivity, respectively. (b) Same as (a) but for the Hall component of the conductivity $\sigma_{xy}$, computed from Eq.~\eqref{eq:sigma_xy_B}. In all cases, we chose $E_{\rm F}=0.2\,$eV and compute the electron damping rate using Eqs.~\eqref{eq:weighted_thermalaverage_scattering} and \eqref{eq:weighted_thermalaverage_scattering_Bfield}.}
\label{fig:conductivity}
\end{center}
\end{figure}

Figure~\ref{fig:temp_dependent_scattering_rate} shows the value of the total electron damping rate as a function of temperature calculated for different magnetic field strengths using Eqs.~\eqref{eq:weighted_thermalaverage_scattering} and \eqref{eq:weighted_thermalaverage_scattering_Bfield}. We assume a DC mobility $\mu_{\rm DC}=10^4\,$cm$^2$/Vs and consider two different situations. In Fig.~\ref{fig:temp_dependent_scattering_rate}(a), the lattice temperature is taken to be equal to the electron temperature (i.e., $T_l=T_e$). In contrast, in Fig.~\ref{fig:temp_dependent_scattering_rate}(b), we assume that the lattice temperature is kept fixed at $T_l=300\,$K, corresponding to a scenario where an intense ultrafast optical pulse heats the structure, leading to a picosecond-scale transient during which the electron and lattice temperatures are completely decoupled \cite{dias2020thermal}. As expected, in all of the cases under consideration, $\tau^{-1}$ increases with temperature. Comparing the results at $1200\,$K with the value of the electron damping rate used in the main text, we find that when $T_l=T_e$, $\tau^{-1}$ increases by a factor of $\sim 5$ for ${E_{\rm F}}=0.2\,$eV and by a factor of $\sim 16$ for ${E_{\rm F}}=0.4\,$eV. These factors decrease to $\sim 2$ and $\sim 4$, respectively, when $T_l$ is held constant at $300\,$K.

Incidentally, the presence of a static magnetic field appears to have a negligible impact on the electron damping rate. However, it is important to note that Ref.~\cite{sohier2014phonon}, from which we obtain the expressions and parameters for the temperature-dependent electron damping rate, does not account for a magnetic field in their DFT calculations, which might influence the results.  

Figures~\ref{fig:conductivity}(a) and (b) present the longitudinal and Hall components of the conductivity for different temperatures and magnetic fields, as indicated by the labels. We assume $E_{\rm F}=0.2\,$eV and compute the electron damping rate using Eqs.~\eqref{eq:weighted_thermalaverage_scattering} and \eqref{eq:weighted_thermalaverage_scattering_Bfield}. The real and imaginary parts of the different conductivity components are shown with solid and dashed curves, respectively.  At zero temperature, the effect of interband transitions becomes visible at $\hbar\omega=2E_{\rm F}$ for $B=0$, while for $B\neq 0$, we also observe Shubnikov-de Haas oscillations \cite{ferreira2011faraday}.  As the temperature increases the transitions between LL become thermally accessible at lower energies, giving rise to the hybridization effects discussed in the main text.

\section{Extinction Cross-Section of Magneto-Optical Graphene Nanodisks and Comparison with Finite Element Method Simulations}

We characterize the optical response of an individual graphene nanodisk by computing its extinction cross-section. Assuming excitation by a plane wave propagating along $-\hat{\bf z}$ with electric field polarization $\hat{\bf e}$, the extinction cross-section is calculated as \cite{hohenester2020nano}
\begin{equation} \label{eq:sigma_ext_def}
\sigma^{\rm ext}= \frac{4 \pi k}{\sqrt{\epsilon}} {\rm Im}\{ \hat{\bf e}^* \cdot \alpha \cdot \hat{\bf e} \} .
\end{equation}
where $\epsilon$ is the dielectric function of the medium around the nanodisk. When excited with circularly polarized light, defined by the polarization vector $\hat{\bf e}_\pm=(\hat{\bf x}\pm {\rm i}\hat{\bf y})/\sqrt{2}$, where the positive and negative sign correspond to LCP and RCP light, respectively, Eq.~\eqref{eq:sigma_ext_def} reduces to
\begin{equation}
\sigma_{\pm}^{\rm ext} = \frac{4 \pi k}{\sqrt{\epsilon}} \left( {\rm Im}\{\alpha_{xx}\}\pm{\rm Re}\{\alpha_{xy}\} \right).  \nonumber \\
\end{equation}

\begin{figure} 
\begin{center}
\includegraphics[width=0.431\linewidth]{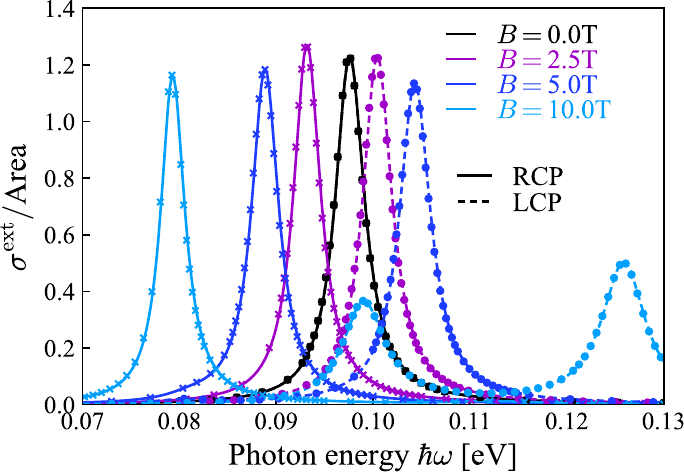}
\caption{{\bf Benchmark of the semianalytical approach against fully numerical calculations.} Extinction cross-section spectra for a individual graphene nanodisk in vacuum. The nanodisk has $D=120\,$nm, $E_{\rm F}=0.2\,$eV and $T=300\,$K.  We consider four different magnetic field strengths, as indicated in the legend. Dashed and solid curves represent the results for the semianalytical model under LCP and RCP excitation, while dots and crosses indicate the corresponding results obtained from direct numerical solutions of Maxwell’s equations obtained using a FEM.}
\label{fig:ext_Comsol}
\end{center}
\end{figure}

In Fig.~\ref{fig:ext_Comsol} we analyze the extinction cross-section of a graphene nanodisk in vacuum for different magnetic field strengths, as indicated by the legend. We consider a nanodisk with $D=120\,$nm, $E_{\rm F}=0.2\,$eV, and $T=300\,$K. Dashed and solid curves represent the results obtained with the semianalytical model developed in this work for LCP and RCP light, respectively. These results are in excellent agreement with direct numerical solutions of Maxwell's equations obtained using the finite element method (FEM), which are displayed with circles and crosses for LCP and RCP light. The FEM calculations are performed using the commercial software COMSOL Multiphysics. Following previous works \cite{deop2022optical}, we model the graphene nanodisk by defining a two-dimensional circular boundary with a surface current boundary condition described by the tensorial Ohm’s law $J_i({\bf R},\omega) =\sigma_{ij}(\omega)E_j({\bf R},\omega)$. We employ a spherical domain, which is truncated by means of a scattering boundary condition. Invoking the optical theorem, we compute the extinction cross-section from the imaginary part of the far-field scattering amplitude in the forward direction.  All of the results are checked for convergence with respect to mesh and domain size.
It is worth noting that both approaches capture the splitting associated with the thermal magnetoplasmon (TMP) resulting from the hybridization with the $n=0\to n=1$ LL transition.

\begin{figure} 
\begin{center}
\includegraphics[width=0.497\linewidth]{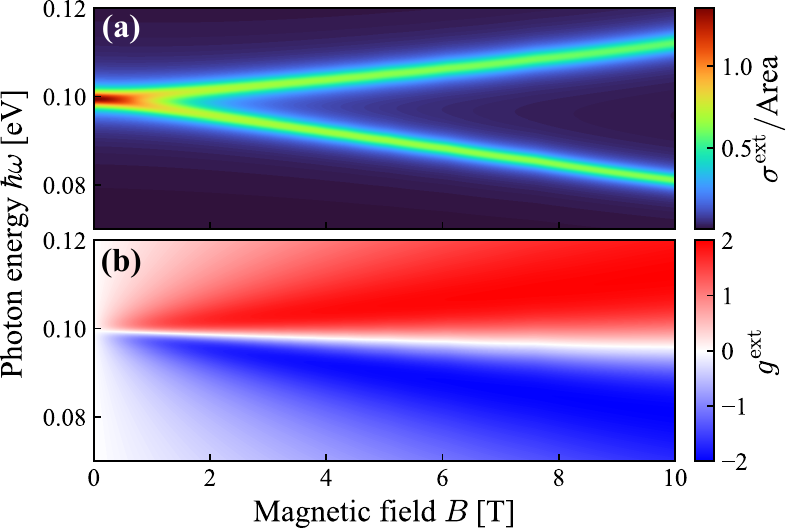}
\caption{{\bf Magneto-optical graphene nanodisk at zero temperature.} (a) Extinction cross-section of a graphene nanodisk in vacuum for linearly polarized light as a function of $B$. We assume $D=120\,$nm and $E_{\rm F}=0.2\,$eV. (b) Extinction dissymmetry factor corresponding to the results of panel (a), calculated using Eq.~\eqref{eq:extinction_dissymmetry_factor} of the main text.}
\label{fig:zero_temp}
\end{center}
\end{figure}

As a reference, Fig.~\ref{fig:zero_temp} presents the extinction cross-section and the extinction dissymmetry factor for an individual graphene nanodisk in vacuum at $T=0\,$K, plotted as a function of the magnetic field strength. We assume the same parameters as in Fig.~\ref{fig:fig1} of the main text, namely, $D=120\,$nm and $E_{\rm F}=0.2\,$eV.  The cross-section spectrum of Fig.~\ref{fig:zero_temp}(a) displays the expected magnetoplasmon splitting for linearly-polarized excitation, with the branches at high and low frequencies being excited, respectively, by the LCP and RCP components of the incoming light. The analysis of $g^{\rm ext}$, shown in Fig.~\ref{fig:zero_temp}(b), confirms the strong chiral character of the magnetoplasmon supported by the graphene nanodisk.

\begin{figure} 
\begin{center}
\includegraphics[width=0.614\linewidth]{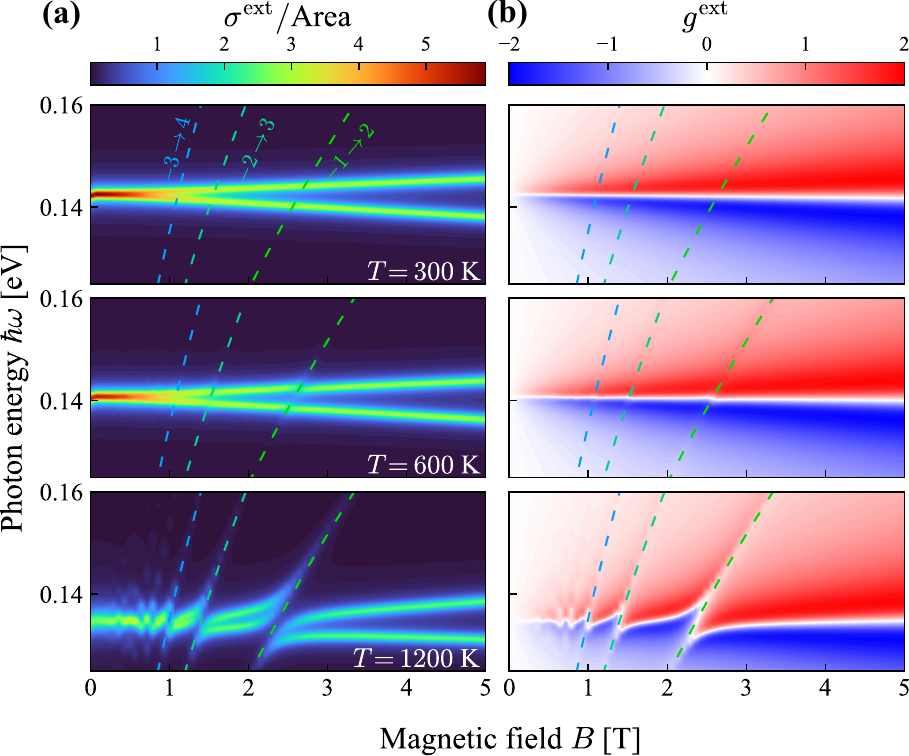}
\caption{{\bf Magneto-optical graphene nanodisk and thermal magnetoplasmons.}  (a) Extinction cross-section of a graphene nanodisk in vacuum for linearly polarized light as a function of $B$. We assume $D=120\,$nm, $E_{\rm F}=0.4\,$eV, and consider three different temperatures: $T=300\,$K, $T=600\,$K, and $T=1200\,$K. (b) Extinction dissymmetry factor corresponding to the results of panel (a), calculated using Eq.~\eqref{eq:extinction_dissymmetry_factor} of the main text. The dashed lines in both panels signal the transitions between LLs and are color-coded to match the arrows in Fig.~\ref{fig:fig1}(b) of the main text.}\label{fig:ext_dissymmetry_factor}
\end{center}
\end{figure}

\begin{figure}[h!]
\begin{center}
\includegraphics[width=0.853\linewidth]{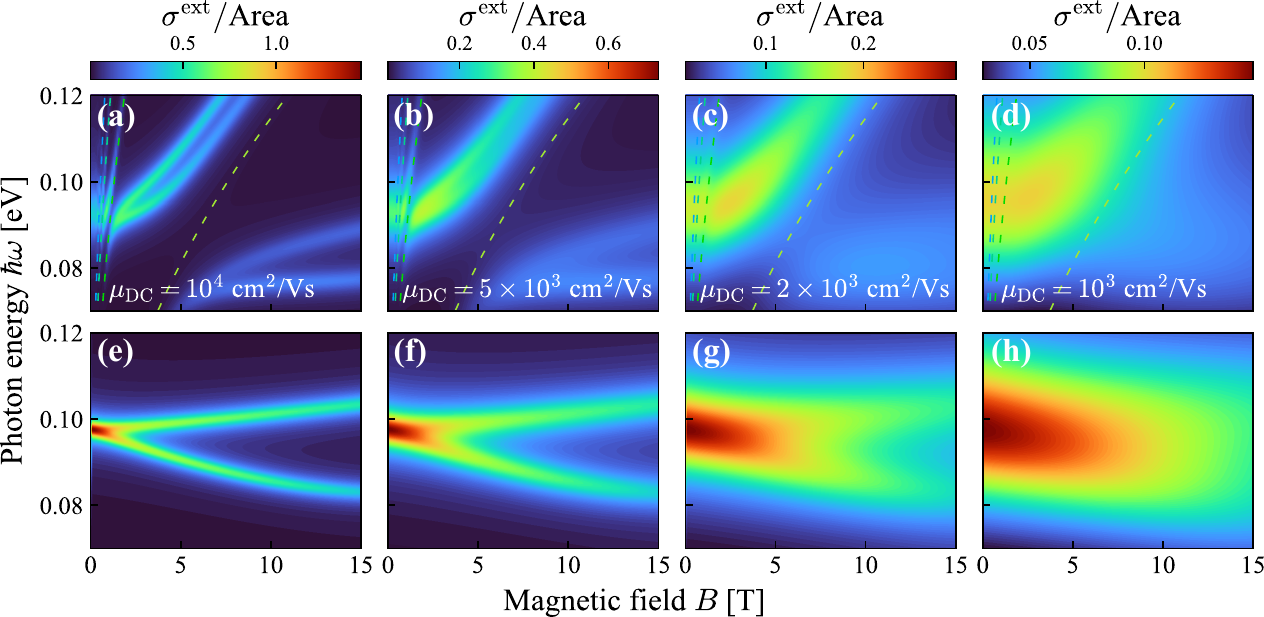}
\caption{{\bf Effect of electron damping rate on the magneto-optical response of a graphene nanodisk.}  (a-d) Extinction cross-section of a graphene nanodisk in vacuum for linearly polarized light as a function of $B$. We assume $D=120\,$nm, $E_{\rm F}=0.2\,$eV, and $T=1200\,$K, while varying $\mu_{\rm DC}$ across the panels, as indicated by the labels. This variation results in different electron damping rates according to $\tau^{-1} = e v_{\rm F}^2 / \mu_{\rm DC} E_{\rm F}$. Note that panel (a) coincides with the bottom plot of Fig.~\ref{fig:fig1}(d) of the main text.  (e-h) Same as (a-d) but with the contributions of all interband LL transitions, as well as the $n=0\to n=1$ and $n=-1\to n=0$ transitions, artificially removed from Eqs.~\eqref{eq:sigma_xx_B} and \eqref{eq:sigma_xy_B}. }
\label{fig:varying_loss}
\end{center}
\end{figure}

To complement the analysis of Figs.~\ref{fig:fig1}(d) and (e) of the main text, Fig.~\ref{fig:ext_dissymmetry_factor} displays similar results for a larger carrier concentration. Specifically,  Figs.~\ref{fig:ext_dissymmetry_factor}(a) and (b) show the extinction cross-section and the corresponding extinction dissymmetry factor for a graphene nanodisk with $D=120\,$nm and $E_{\rm F}=0.4\,$eV as a function of $B$, across three different temperatures. From these results, we observe that increasing the carrier concentration from $E_{\rm F}=0.2\,$eV to $E_{\rm F}=0.4\,$eV results in a shift of the magnetoplasmon to higher energies. This, in turn, drives the TMPs to appear at larger values of the magnetic field. Consequently, only the TMPs associated with the interband LL transitions are visible, as the $n=0\to n=1$ LL transition lies beyond of the range of $B$ under consideration. Additionally, the emergence of strong anti-crossing features associated with the TMPs requires higher temperatures than those of Fig.~\ref{fig:fig1} of the main text, owing to the increased occupation of the LLs involved in these transitions.

Figures~\ref{fig:varying_loss}(a)-(d) illustrate the effect of the electron damping rate on the magneto-optical response of a graphene nanodisk. Similar to the results of Fig.~\ref{fig:fig1}(d), we calculate the extinction cross-section of a graphene nanodisk in vacuum for linearly polarized light as a function of $B$, considering decreasing values of the DC mobility $\mu_{\rm DC}$. Since, we assume that $\tau^{-1} = e v_{\rm F}^2 / \mu_{\rm DC} E_{\rm F}$ in the main text, reducing $\mu_{\rm DC}$ is equivalent to increasing the electron damping rate. We choose a nanodisk with $D=120\,$nm, $E_{\rm F}=0.2\,$eV, $T=1200\,$K, and decrease $\mu_{\rm DC}$ from the value used in the main text $10^4\,$cm$^2$/Vs to $10^3\,$cm$^2$/Vs. Examining the results, it is evident that the increase of $\tau^{-1}$ smears out the different spectral features. However, the self-hybridization of the TMPs is still clearly visible for $\mu_{\rm DC} = 2000\,$cm$^2$/Vs. Note that the value of $\tau$ corresponding to the mobility of Fig.~\ref{fig:varying_loss}(b) is approximately equal to the temperature-dependent $\tau$ at $T=1200\,$K for the mobility used in the main text, when $T_l$ is kept fixed at $300\,$K.

To further investigate the effect of increasing the electron damping rate, we repeat the calculations of Fig.~\ref{fig:varying_loss}(a)-(d) but with the contributions of all interband LL transitions, as well as the $n=0\rightarrow n=1$ and $n=-1\rightarrow n=0$ transitions, artificially removed. Comparing the results, which are shown in Fig.~\ref{fig:varying_loss}(e)-(h), with the original calculations, we clearly observe the impact of the $n=0\rightarrow n=1$ transition, which bends the TMP resonances even for the smallest mobility under consideration. For this mobility, the splitting induced by the magnetic field is no longer discernible but the bending of the TMP resonances resulting from the avoided crossing with the $n=0\rightarrow n=1$ LL transition is still pronounced.

\section{Coupled Dipole Model for Periodic Arrays}

We model the optical response of a periodic array of magneto-optical graphene nanodisks using the coupled dipole model \cite{zhao2003extinction,garciadeabajo2007colloquium}. This approach, which is based on the dipolar approximation, provides very accurate results when the nanostructures are small compared to the wavelength of light and the period of the array \cite{cuartero2020super,zundel2021lattice}. We consider a square array of period $a$, placed in the $z=0$ plane, and surrounded by a homogeneous medium with dielectric function $\epsilon$. Within this approach, each nanodisk is described as a point dipole ${\bf p}_i$ with components along $\hat{\bf x}$ and $\hat{\bf y}$. This dipole satisfies ${\bf p}_i=\alpha\cdot{\bf E}({\bf R}_i)$, where $ \alpha$ is the tensorial polarizability of the nanodisk and ${\bf E}$ is the total self-consistent electric field acting on it. The latter is the sum of the external electric field, and the electric field produced by all of the other elements in the array
\begin{equation}\label{eq:pi}
{\bf p}_i = \alpha \cdot {\bf E}^{\rm ext}({\bf R}_i) + \alpha \cdot \sum_{j\neq i} {\bf G}_{ij} \cdot {\bf p}_j,
\end{equation}
In this expression, ${\bf R}_i$ is the position of the $i^{\rm th}$ nanodisk and ${\bf G}_{ij}$ is the dyadic Green tensor for a homogeneous medium with dielectric function $\epsilon$, which is defined as \cite{novotny2012principles, zundel2022lattice}:
\begin{equation}
{\bf G}_{ij} = \left[k^2\epsilon\mathbbm{1}_{2\times2} + \nabla_{\bf R} \otimes\nabla_{\bf R} \right]\frac{{\rm e}^{{\rm i}k\sqrt{\epsilon}|{\bf R}_i - {\bf R}_j|}}{\epsilon|{\bf R}_i - {\bf R}_j|}. \nonumber
\end{equation}
The external field acting on the array is ${\bf E}^{\rm ext}({\bf R}_i)={\bf E}^{\rm ext} {\rm e}^{{\rm i} {\bf k}_{\parallel} \cdot {\bf R}_i}$, where ${\bf k}_{\parallel}$ is the component of the wave vector parallel to the array. The periodic nature of the array guarantees a solution of Eq.~\eqref{eq:pi} in the form of Bloch waves ${\bf p}_i = {\bf p} {\rm e}^{{\rm i} {\bf k}_{\parallel} \cdot {\bf R}_i}$, where
\begin{equation}
{\bf p} = \alpha^{\rm eff} \cdot {\bf E}^{\rm ext} \nonumber 
\end{equation}
and
\begin{equation}
\alpha^{\rm eff} = \left[\alpha^{-1}  -  \mathcal{G}  \right]^{-1} \nonumber
\end{equation}
is the effective polarizability of the array. In the last expression, we have introduced the lattice sum $\mathcal{G} = \sum_{j\neq 0} {\bf G}_{j0}{\rm e}^{-{\rm i} {\bf k}_{\parallel} \cdot {\bf R}_j}$, which is an entirely geometrical parameter that depends on the specific form of the array. To compute it, we use Ewald's method \cite{jordan1986efficient, kolkowski2019lattice, zundel2021lattice}.

Once the dipole induced in the nanodisks of the array is known, we can calculate the electric field scattered by them at a point ${\bf r} = {\bf R} + z \hat{\bf z}$ as \cite{zundel2021lattice}
\begin{equation}
{\bf E}^{\rm sc}({\bf r})=  \sum_{i} \left[k^2\epsilon  {\bf p}_{i}  + ({\bf p}_{i} \cdot \nabla_{\bf r}) \nabla_{\bf r} \right] \frac{{\rm e}^{ {\rm i} k\sqrt{\epsilon}|{\bf r} - {\bf R}_i|}}{\epsilon|{\bf r} - {\bf R}_i|}. \nonumber
\end{equation}
Then, using the Weyl identity \cite{novotny2012principles}, this expression can be rewritten as
\begin{equation}
{\bf E}^{\rm sc}({\bf r})=\sum_{i} \left[k^2\epsilon\, {\bf p} + ({\bf p}\cdot \nabla_{\bf r}) \nabla_{\bf r} \right]  \frac{{\rm i}}{2\pi \epsilon}\int \frac{{\rm d}^2{\bf k}'_{\parallel}}{k'_z} {\rm e}^{{\rm i} {\bf k}'_{\parallel}\cdot(\mathbf{R}-\mathbf{R}_i)} {\rm e}^{{\rm i}k'_z|z|} {\rm e}^{{\rm i} {\bf k}_{\parallel} \cdot {\bf R}_i}, \nonumber
\end{equation}
with $k_z = \sqrt{\epsilon k^2 - |{\bf k}_{\parallel}|^2}$. Exploiting the properties of periodic arrays, the expression above reduces to 
\begin{equation}
{\bf E}^{\rm sc}({\bf r})= \frac{2\pi {\rm i}}{a^2 \epsilon} \sum_{{\bf q}} \left\{k^2\epsilon\,  {\bf p} -  [{\bf p}\cdot({\bf k}+{\bf q})] ({\bf k}+{\bf q})  \right\}{\rm e}^{{\rm i} ({\bf k}_{\parallel}+{\bf q})\cdot \mathbf{R} } \frac{{\rm e}^{{\rm i}k_{{\bf q},z}|z|}}{k_{{\bf q},z}} , \nonumber
\end{equation}
where ${\bf q}$ are the vectors of the reciprocal lattice and $k_{{\bf q},z} = \sqrt{\epsilon k^2 - |({\bf k}_{\parallel}+{\bf q})|^2}$. When the period of the array is smaller than the wavelength of light, the only diffraction order in the sum that is not evanescent is the zeroth order. Then, taking into account that in this work we focus on normal incidence excitation and neglecting all of the evanescent components, the field scattered by the array reduces to
\begin{equation}
{\bf E}^{\rm sc}({\bf r})= \frac{2\pi {\rm i} k}{a^2 \sqrt{\epsilon}} \alpha^{\rm eff}\cdot {\bf E}^{\rm ext} {\rm e}^{{\rm i}k_{z}|z|}, \nonumber
\end{equation}
Therefore, within this approximation, we can describe the optical response of the array of nanodisks using the following matrix of reflection coefficients
\begin{equation}\label{eq:reflection}
\begin{pmatrix}
r^{\rm Array}_{xx} & r^{\rm Array}_{xy} \\ r^{\rm Array}_{yx} & r^{\rm Array}_{yy}
\end{pmatrix}= \frac{2\pi{\rm i} k}{a^2\sqrt{\epsilon}}  \alpha^{\rm eff}. 
\end{equation}

\section{Transfer Matrix Method for Stratified Media}

To calculate the optical response of the nanodisk array combined with a dielectric spacer and mirror in a Salisbury screen configuration, we use a transfer matrix method for stratified media. Specifically, considering circularly polarized light that propagates perpendicular to the system, we can decompose the electric field as \cite{ferreira2011faraday}
\begin{equation}
E_{n,\nu}^{\uparrow\downarrow} = E_{n,x}^{\uparrow\downarrow} - {\rm i} \nu E_{n,y}^{\uparrow\downarrow}, \nonumber
\end{equation}
where $n$ labels the different media, $\uparrow$ ($\downarrow$) indicates that the field propagates along $\hat{\bf z}$ (-$\hat{\bf z}$), and $\nu = \pm$ denotes the polarization of the field. Importantly, we define the handedness such that for a field propagating along $-\hat{\bf z}$, $+$ and $-$ corresponds to LCP and RCP, respectively, while the definition is reverted when the field propagates along $\hat{\bf z}$.

Following the usual transfer matrix approach, the fields at the interface between two isotropic nonmagnetic media indexed by $n$ and $m$ are related by \cite{hohenester2020nano,ferreira2011faraday}:
\begin{equation}
    \begin{pmatrix}
        E_{n,\nu}^{\uparrow} \\
        E_{n,\nu}^{\downarrow}
    \end{pmatrix} 
    =
    {\bf N}_{\nu}^{n,m}\cdot     
    \begin{pmatrix}
        E_{m,\nu}^{\uparrow} \\
        E_{m,\nu}^{\downarrow}
    \end{pmatrix}, \nonumber
\end{equation}
where 
\begin{equation}
     {\bf N}_{\nu}^{nm} = \frac{1}{2k_{n}} \begin{pmatrix}
        k_{n} + k_{m}& k_{n} - k_{m} \\
       k_{n} - k_{m} & k_{n} + k_{m}
    \end{pmatrix}, \nonumber
\end{equation}
and $k_{n} = \sqrt{\epsilon_n}k$. The propagation through an isotropic nonmagnetic media of thickness $d_n$ is described by the matrix
\begin{equation}
     {\bf P}^n = \begin{pmatrix}
        {\rm e}^{{\rm i} k_{n} d_n} &0 \\ 0 & {\rm e}^{-{\rm i} k_{n} d_n}
    \end{pmatrix}. \nonumber
\end{equation}
Regarding the array of graphene nanodisks, we follow the procedure of Ref.~\cite{thongrattanasiri2012complete} and find the corresponding transfer matrix by using Eq.~\eqref{eq:reflection} and considering incoming fields from above and below the array. This results in the following equations:
\begin{gather}
    \begin{pmatrix}
        E_{{\rm a},x}^{\uparrow}\\ E_{{\rm a},y}^{\uparrow}
    \end{pmatrix} = \frac{2\pi{\rm i} k}{a^2\sqrt{\epsilon}} \alpha^{\rm eff} \cdot \begin{pmatrix}
        E_{{\rm a},x}^{\downarrow} \\ E_{{\rm a},y}^{\downarrow}
    \end{pmatrix}, \qquad
    \begin{pmatrix}
        E_{{\rm b},x}^{\downarrow}\\ E_{{\rm b},y}^{\downarrow}
    \end{pmatrix} = \left( \mathbbm{1}_{2\times2}+\frac{2\pi{\rm i} k}{a^2\sqrt{\epsilon}}  \alpha^{\rm eff}\right) \cdot \begin{pmatrix}
        E_{{\rm a},x}^{\downarrow} \\ E_{{\rm a},y}^{\downarrow}
    \end{pmatrix}, \nonumber \\
    \begin{pmatrix}
        E_{{\rm b},x}^{\downarrow}\\ E_{{\rm b},y}^{\downarrow}
    \end{pmatrix} = \frac{2\pi{\rm i} k}{a^2\sqrt{\epsilon}}  \alpha^{\rm eff} \cdot \begin{pmatrix}
        E_{{\rm b},x}^{\uparrow} \\ E_{{\rm b},y}^{\uparrow}
    \end{pmatrix}, \qquad
    \begin{pmatrix}
        E_{{\rm a},x}^{\uparrow}\\ E_{{\rm a},y}^{\uparrow}
    \end{pmatrix} = \left( \mathbbm{1}_{2\times2}+\frac{2\pi{\rm i} k}{a^2\sqrt{\epsilon}} \alpha^{\rm eff}\right) \cdot \begin{pmatrix}
        E_{{\rm b},x}^{\uparrow} \\ E_{{\rm b},y}^{\uparrow}
    \end{pmatrix}.    \nonumber
\end{gather}
Noting that $\alpha^{\rm eff}_{yx}=-\alpha^{\rm eff}_{xy}$, the transfer matrix for the array of nanodisks is then given by
\begin{equation}
   {\bf N}_{\nu}^{\rm Array} =\frac{1}{1+K\alpha^{\rm eff}_{\nu}} \begin{pmatrix}
        1+2K\alpha^{\rm eff}_{\nu} & K\alpha^{\rm eff}_{\nu}\\
        -K\alpha^{\rm eff}_{\nu} & 1 
    \end{pmatrix}\nonumber
\end{equation}
with $K = 2\pi {\rm i} k / a^2\sqrt{\epsilon}$ and $\alpha^{\rm eff}_{\nu} = \alpha^{\rm eff}_{xx}-{\rm i}\nu \alpha^{\rm eff}_{yx}$.

\begin{figure} 
\begin{center}
\includegraphics[width=0.708\linewidth]{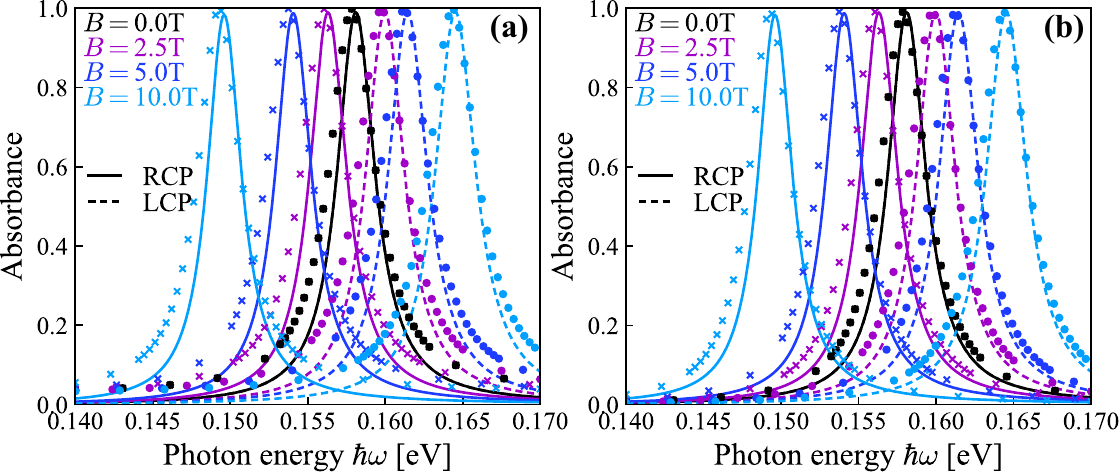}
\caption{{\bf Benchmark of the semianalytical approach against fully numerical calculations.} (a) Absorbance of the system sketched in Fig.~\ref{fig:fig2}(a) the main text, calculated for different magnetic field strengths indicated in the legend. The parameters of the system are as follows: $a=2D$, $D=60\,$nm, $E_{\rm F}=0.4\,$eV, $T=300\,$K, and $d=1\,$$\mu$m. Dashed and solid curves represent the results obtained using the combination of the coupled dipole model and the transfer matrix method under LCP and RCP light, respectively. These results are benchmarked against direct numerical solutions of Maxwell’s equations obtained using a FEM, which are displayed with dots and crosses for LCP and RCP light, respectively. 
(b) Same as in (a) but, in this case, the FEM calculations consider only the absorbance occurring in the graphene nanodisks.}
 \label{fig:Salisbury_Screen_Comsol_comparison}
 \end{center}
\end{figure}

With all of these definitions, we can write the field above the system composed by the periodic array of nanodisks, the dielectric spacer, and the gold mirror, as
\begin{equation}\label{eq:Generalized_transfer_matrix_definition}
    \begin{pmatrix}
        E_{\rm a,\nu}^{\uparrow} \\
        E_{\rm a,\nu}^{\downarrow}
    \end{pmatrix} 
    =
    {\bf N}_{\nu}^{\rm Array} \cdot {\bf N}_{\nu}^{\rm a, Spacer} \cdot {\bf P}^{\rm Spacer} \cdot {\bf N}_{\nu}^{\rm Spacer, Au}\cdot
    \begin{pmatrix}
        E_{\rm Au,\nu}^{\uparrow} \\
        E_{\rm Au,\nu}^{\downarrow}
    \end{pmatrix} 
    =
    {\bf N}_{\nu} \cdot   
    \begin{pmatrix}
        E_{\rm Au,\nu}^{\uparrow} \\
        E_{\rm Au,\nu}^{\downarrow} 
    \end{pmatrix}. \nonumber
\end{equation}
in terms of the field inside the gold mirror. The reflection and transmission coefficients can be obtained by taking ratios between the incoming and outgoing field as
\begin{gather}
r_{\nu} = \frac{E_{\rm a,\nu}^{\uparrow}}{E_{\rm a,\nu}^{\downarrow}} = \frac{\left[{\bf N}_{\nu} \right]_{\uparrow,\downarrow}}{\left[{\bf N}_{\nu} \right]_{\downarrow,\downarrow}}, \nonumber \\
t_{\nu} = \frac{E_{\rm Au,\nu}^{\downarrow}}{E_{\rm a,\nu}^{\downarrow}} = \frac{1}{\left[{\bf N}_{\nu} \right]_{\downarrow,\downarrow}}. \nonumber 
\end{gather}
Using these coefficients, we can write the reflectance as  $\mathcal{R}_{\nu} = |r_{\nu}|^2$, while the absorbance of the array of nanodisks is given by 
\begin{equation}\label{eq:absorbance}
\mathcal{A}_{\nu} = 1 - \mathcal{R}_{\nu} - \frac{ {\rm Re}\left\{ k_{\rm Au}\right\}}{k_{\rm a}} |t_{\nu}|^2. 
\end{equation}

The approach based on the combination of the coupled dipole model and the transfer matrix method described above neglects the contribution of the evanescent components scattered by the array of nanodisks. Furthermore, when calculating the response of the array and, in particular, $\alpha^{\rm eff}$, we assume that the array is embedded in a homogeneous medium with a dielectric function $\epsilon_{\rm eff}$. These approximations allow us to maintain a semianalytical derivation, while still yielding highly accurate results, as we demonstrate in Fig.~\ref{fig:Salisbury_Screen_Comsol_comparison}(a). There, we show the absorbance of the system analyzed in Fig.~\ref{fig:fig2} of the main text. Specifically, the array has periodicity $a=2D$ and is made of nanodisks with diameter $D=60\,$nm, Fermi energy $E_{\rm F}=0.4\,$eV, and temperature of $T=300\,$K. The dielectric spacer has a thickness of $d=1\,$$\mu$m and a dielectric constant of $\epsilon_{\rm d}=2.1$, while for the gold mirror we use a Drude model fitted to the tabulated data in Ref.~\cite{johnson1972optical}. 

The results obtained with Eq.~\eqref{eq:absorbance} are displayed with dashed and solid curves for LCP and RCP light, respectively. These results are benchmarked against direct numerical solutions of Maxwell’s equations obtained using a commercial FEM solver (COMSOL Multiphysics), which are indicated with dots (LCP) and crosses (RCP). For the FEM calculations, we employ a domain corresponding to one unit cell of the system, together with Floquet-periodic boundary conditions. We truncate the cell along $\hat{\bf z}$ by employing perfectly matched layers. As described previously for the extinction cross-section calculations, we model the graphene disks by implementing a surface current boundary condition with the appropriate conductivity. The absorbance from the graphene disks and the gold mirror is calculated separately by integration of the Joule dissipation in their respective domains. All of the results are checked for convergence with respect to mesh and domain size.
 
Upon comparing the results of Eq.~\eqref{eq:absorbance} with the numerical calculations shown in Fig.~\ref{fig:Salisbury_Screen_Comsol_comparison}(a), we observe excellent agreement around the resonance peaks, though it slightly worsens away from them. The origin of this discrepancy is that Eq.~\eqref{eq:absorbance} only considers the absorbance in the array of nanodisks, as the gold mirror has an almost perfect reflectance. However, the evanescent components, which are neglected in the semianalytical approach but captured by the FEM calculations, lead to some absorption in the mirror. To test this hypothesis, Fig.~\ref{fig:Salisbury_Screen_Comsol_comparison}(b) benchmarks the semianalytical results against FEM calculations that, in this case, only consider absorption within the array of graphene nanodisks. The much-improved agreement outside the resonances clearly supports our explanation.


%

\end{document}